\documentclass{elsarticle}

\usepackage[reviewcopy]{adndt}
\usepackage{longtable}

\usepackage{amsmath}

\usepackage{amssymb}

\biboptions{square,sort&compress}
\bibpunct[]{[}{]}{,}{n}{}{;}
\begin{document}

\begin{frontmatter}

\journal{Atomic Data and Nuclear Data Tables}


\title{Discovery of the thallium, lead, bismuth, and polonium isotopes}

\author{C. Fry}
\author{M. Thoennessen\corref{cor1}}\ead{thoennessen@nscl.msu.edu}

 \cortext[cor1]{Corresponding author.}

 \address{National Superconducting Cyclotron Laboratory and \\ Department of Physics and Astronomy, Michigan State University, \\ East Lansing, MI 48824, USA}

\begin{abstract}
Currently, forty-two thallium, forty-two lead, forty-one bismuth, and forty-two polonium isotopes have so far been observed; the discovery of these isotopes is discussed. For each isotope a brief summary of the first refereed publication, including the production and identification method, is presented.
\end{abstract}

\end{frontmatter}





\newpage
\tableofcontents
\listofDtables

\vskip5pc

\section{Introduction}\label{s:intro}

The discovery of thallium, lead, bismuth, and polonium isotopes is discussed as part of the series summarizing the discovery of isotopes, beginning with the cerium isotopes in 2009 \cite{2009Gin01}. Guidelines for assigning credit for discovery are (1) clear identification, either through decay-curves and relationships to other known isotopes, particle or $\gamma$-ray spectra, or unique mass and Z-identification, and (2) publication of the discovery in a refereed journal. The authors and year of the first publication, the laboratory where the isotopes were produced as well as the production and identification methods are discussed. When appropriate, references to conference proceedings, internal reports, and theses are included. When a discovery includes a half-life measurement the measured value is compared to the currently adopted value taken from the NUBASE evaluation \cite{2003Aud01} which is based on the ENSDF database \cite{2008ENS01}. In cases where the reported half-life differed significantly from the adopted half-life (up to approximately a factor of two), we searched the subsequent literature for indications that the measurement was erroneous. If that was not the case we credited the authors with the discovery in spite of the inaccurate half-life. All reported half-lives inconsistent with the presently adopted half-life for the ground state were compared to isomer half-lives and accepted as discoveries if appropriate following the criterium described above.

The first criterium is not clear cut and in many instances debatable. Within the scope of the present project it is not possible to scrutinize each paper for the accuracy of the experimental data as is done for the discovery of elements \cite{1991IUP01}. In some cases an initial tentative assignment is not specifically confirmed in later papers and the first assignment is tacitly accepted by the community. The readers are encouraged to contact the authors if they disagree with an assignment because they are aware of an earlier paper or if they found evidence that the data of the chosen paper were incorrect.

The discovery of several isotopes has only been reported in conference proceedings which are not accepted according to the second criterium. One example from fragmentation experiments why publications in conference proceedings should not be considered are $^{118}$Tc and $^{120}$Ru which had been reported as being discovered in a conference proceeding \cite{1996Cza01} but not in the subsequent refereed publication \cite{1997Ber01}.

The initial literature search was performed using the databases ENSDF \cite{2008ENS01} and NSR \cite{2008NSR01} of the National Nuclear Data Center at Brookhaven National Laboratory. These databases are complete and reliable back to the early 1960's. For earlier references, several editions of the Table of Isotopes were used \cite{1940Liv01,1944Sea01,1948Sea01,1953Hol02,1958Str01,1967Led01}. A good reference for the discovery of the stable isotopes was the second edition of Aston's book ``Mass Spectra and Isotopes'' \cite{1942Ast01}. For the isotopes of the radioactive decay chains several books and articles were consulted, for example, the 1908 edition of ``Gmelin-Kraut's Handbuch der anorganischen Chemie'' \cite{1908Fri01}, Soddy's 1911 book ``The chemistry of the radio-elements'' \cite{1911Sod01}, the 1913 edition of Rutherford's book ``Radioactive substances and their radiations'' \cite{1913Rut01}, and the 1933 article by Mary Elvira Weeks ``The discovery of the elements. XIX. The radioactive elements'' published in the Journal of Chemical Education \cite{1933Wee01}. In addition, the wikipedia page on the radioactive decay chains was a good starting point \cite{2011wik02}.

The isotopes within the radioactive decay chains were treated differently. Their decay properties were largely measured before the concept of isotopes was established. Thus we gave credit to the first observation and identification of a specific activity, even when it was only later placed properly within in the decay chain.

\begin{figure}
	\centering
	\includegraphics[scale=.9]{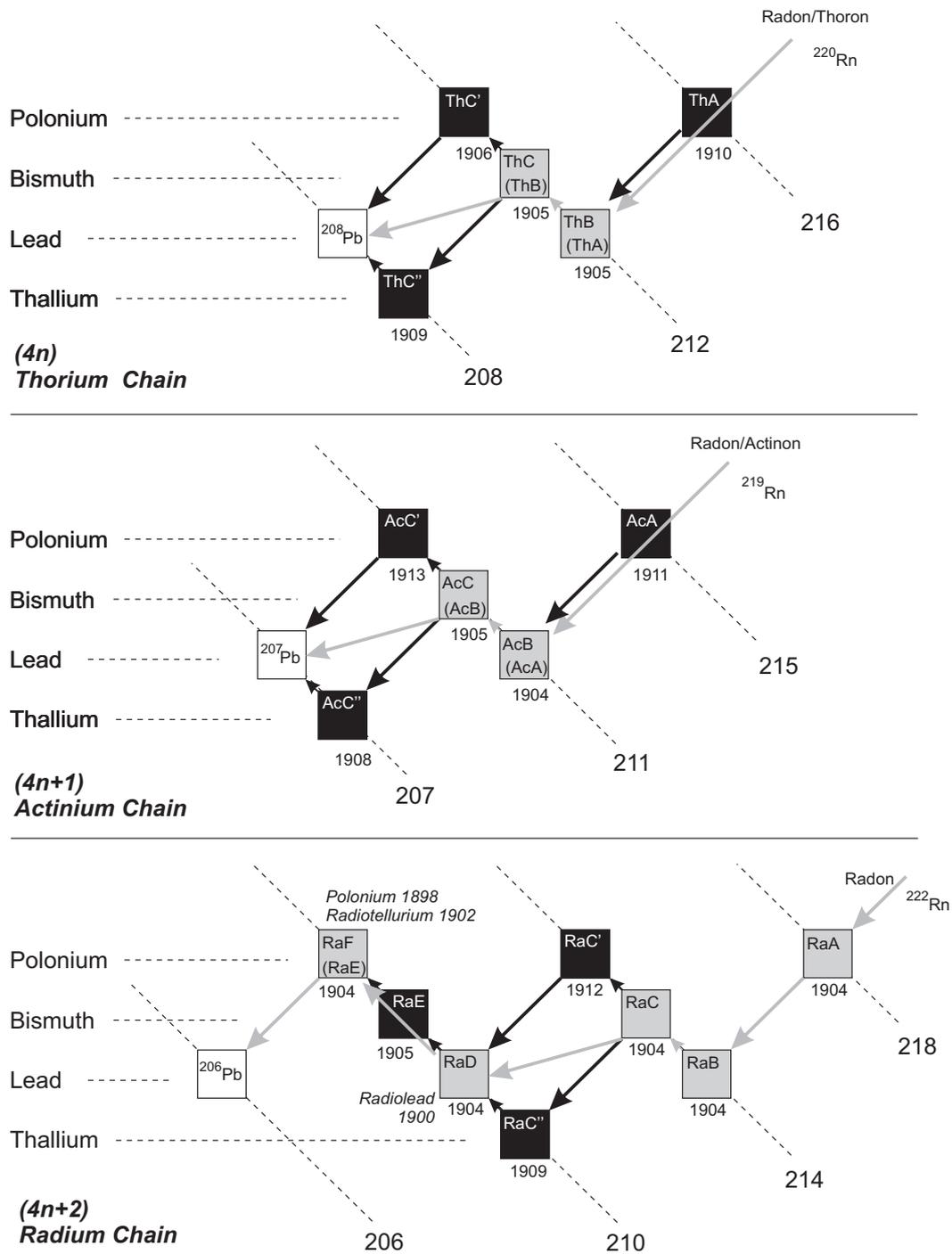}
	\caption{Original nomenclature of thallium, lead, bismuth, and polonium isotopes within the three natural occurring radioactive decay series. The grey squares connected by the grey arrows depict the activities labeled by Rutherford in his Bakerian lecture \cite{1905Rut01}. Names that were changed later are indicated in brackets. The black squares correspond to radioactive substances discovered later.}
\label{f:chain}
\end{figure}

Figure \ref{f:chain} summarizes the isotopes of the three natural ocuring radioactive decay series with their original nomenclature. This notation of the original substances introduced by Rutherford during his Bakerian lecture presented on May 19$^{th}$ 1904 and published a year later \cite{1905Rut01} are shown by grey squares and connected by the grey arrows representing $\alpha$ and $\beta$ decay. Some had to be renamed later (Rutherford's label are listed in brackets) when previously termed ``complex'' activities were separated into two different substances. These are shown as black squares with the corresponding decays shown by black arrows. The white squares show the final stable lead isotopes of the series. Also indicated in the figure are the even earlier names for radium D (radiolead) and radium F (polonium and radio tellurium).





\section{$^{176-217}$Tl}\vspace{0.0cm}

\begin{figure}
	\centering
	\includegraphics[scale=.7]{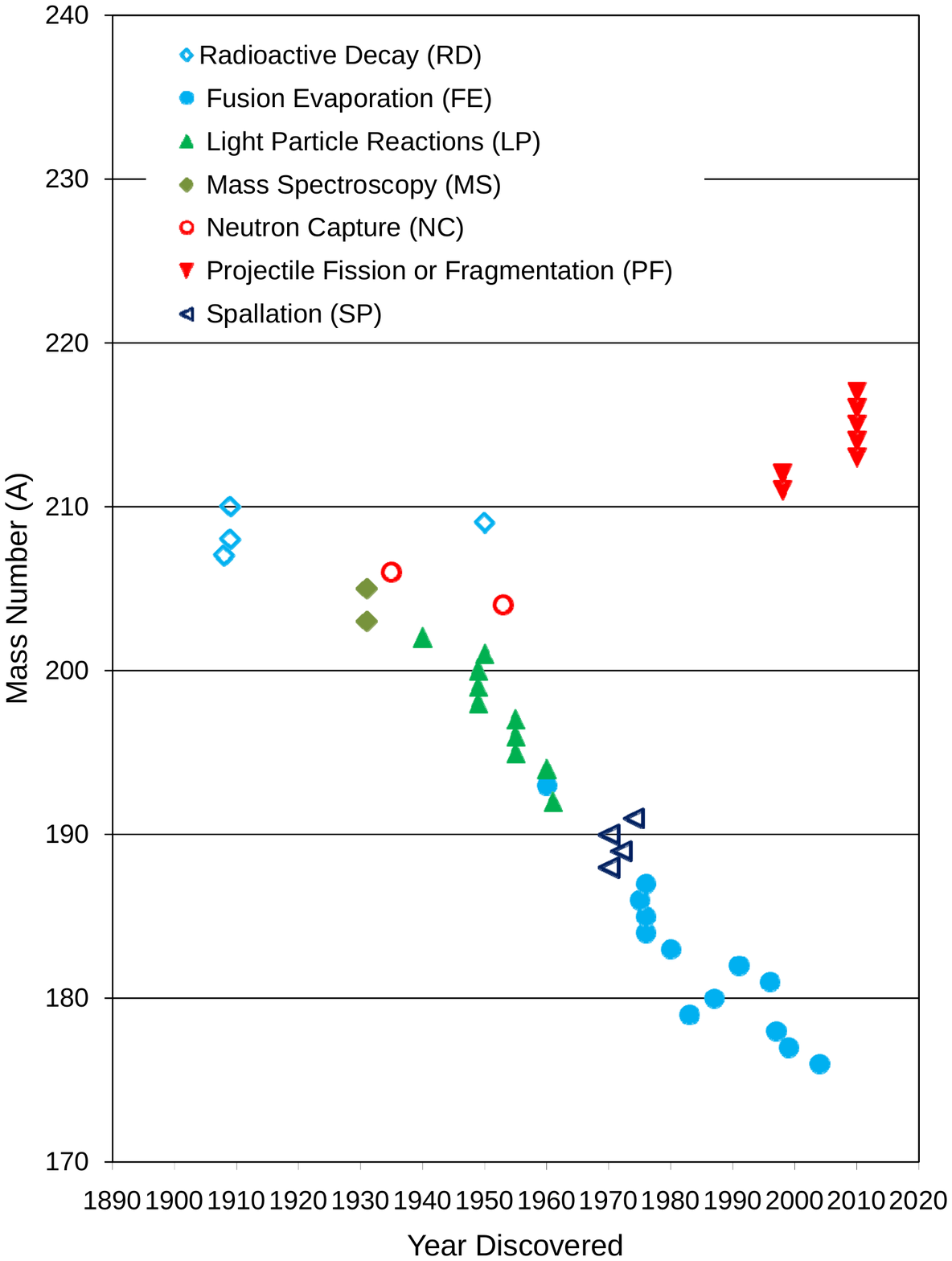}
	\caption{Thallium isotopes as a function of time when they were discovered. The different production methods are indicated.}
\label{f:year-thallium}
\end{figure}

Forty-two thallium isotopes from A = 176--217 have been discovered so far; these include 2 stable, 27 neutron-deficient and 13 neutron-rich isotopes. According to the HFB-14 model \cite{2007Gor01} about 42 additional thallium isotopes could exist. Figure \ref{f:year-thallium} summarizes the year of first discovery for all thallium isotopes identified by the method of discovery: radioactive decay (RD), mass spectroscopy (MS), fusion evaporation reactions (FE), light-particle reactions (LP), neutron capture (NC), projectile fission or fragmentation (PF), and spallation (SP). In the following, the discovery of each thallium isotope is discussed in detail and a summary is presented in Table 1.

\subsection*{$^{176}$Tl}
$^{176}$Tl was discovered by Kettunen et al.\ in 2004, and published in ``Decay studies of $^{170,171}$Au, $^{171-173}$Hg, and $^{176}$Tl'' \cite{2004Ket01}. The Jyv\"askyl\"a K130 cyclotron was used to bombard an enriched $^{102}$Pd target with 380$-$389~MeV $^{78}$Kr beam to produce $^{176}$Tl in the (p3n) fusion-evaporation reaction. $^{176}$Tl was separated with the gas-filled recoil separator RITU, and implanted into a position sensitive silicon strip detector. The identification was based on correlations between the implantation and the subsequent proton emission. ``Based on the properties of eight ER-p$_m$-$\alpha_d$ correlated decay chains in [the figure], the daughter activity of group $^{176}$Tl$^g$ was identified to originate from the $\alpha$ decay of $^{175}$Hg... Based on the properties of the decay chains the mother activity is concluded to originate from the proton emission of $^{176}$Tl to the ground state of $^{175}$Hg.'' The measured half-life of 5.2$^{+3.0}_{-1.4}$~ms is currently the only measurement.

\subsection*{$^{177}$Tl}
Poli et al.\ reported the first observation of $^{177}$Tl in the 1999 paper ``Proton and $\alpha$ radioactivity below the Z = 82 shell closure'' \cite{1999Pol01}. A 370~MeV $^{78}$Kr beam from the Argonne ATLAS accelerator facility bombarded an enriched $^{102}$Pd target and $^{177}$Tl was produced in the (1p2n) fusion-evaporation reaction. Reactions were separated with the Argonne Fragment Mass Analyzer (FMA) and implanted in a double-sided silicon strip detector which also detected subsequent proton and $\alpha$ decay. ``[The figure] shows a composite spectrum of weaker proton and alpha-decay groups [E$_p$ = 1156(20)~keV and E$_\alpha$ = 6907(7)~keV] assigned to the ground-state decays of $^{177}$Tl with branches b$_p$=0.27(13) and b$_\alpha$=0.73(13), and a combined half-life value of 18(5)~ms.'' This is currently the only measurement. A previous search for $^{177}$Tl was not successful \cite{1991Sel01}.

\subsection*{$^{178}$Tl}
In 1997, Carpenter et al.\ observed $^{178}$Tl in ``Excited states in $^{176,178}$Hg and shape coexistence in very neutron-deficient Hg isotopes'' \cite{1997Car01}. A rhodium target was bombarded with 340 and 380~MeV $^{78}$Kr beams from the Argonne ATLAS superconducting linear accelerator forming $^{178}$Tl in the fusion-evaporation reaction $^{103}$Rh($^{78}$Kr,3n). Reactions were separated with the Argonne Fragment Mass Analyzer (FMA) and implanted in a double-sided silicon strip detector which also detected subsequent $\alpha$ decay. ``In addition, three other $\alpha$ lines at 6.71, 6.79, and 6.87 MeV are correlated with A = 178. These three lines are, in turn, all followed by $\alpha$ decay to the same daughter nucleus, $^{174}$Au. They have been associated with the decay of the previously unknown isotope $^{178}$Tl.'' The first measurement of the half-life of 254$^{+11}_{-9}$~ms for $^{178}$Tl was reported five years later confirming the $\alpha$-decay energies \cite{2002Row01}.

\subsection*{$^{179}$Tl}
In the 1983 paper ``Alpha decay of new neutron deficient gold, mercury and thallium isotopes'' Schneider et al.\ reported the discovery of $^{179}$Tl \cite{1983Sch01}. An yttrium target was bombarded with 4.50$-$5.40~MeV/u $^{92}$Mo beams from the GSI UNILAC producing $^{179}$Tl in the (2n) fusion-evaporation reactions. Reaction products were separated with the velocity filter SHIP and implanted in an array of seven position sensitive silicon surface barrier detectors which also measured subsequent $\alpha$ decays. ``The decays of the new isotopes $^{173}$Au, $^{175,176}$Hg, and $^{179}$Tl could be correlated to the known $\alpha$ decays of their daughters.'' The measured half-life of 0.16($^{+9}_{-4}$)~s agrees with the currently accepted value of 0.23(4)~s.

\subsection*{$^{180}$Tl}
$^{180}$Tl was identified in 1987 by Lazarev et al.\ in ``Observation of delayed nuclear fission in the region of $^{180}$Hg'' \cite{1987Laz01}. The Dubna U-400 cyclotron was used to bombard an enriched $^{144}$Sm target with a 230 MeV $^{40}$Ca beam. $^{180}$Tl was produced in the fusion evaporation reaction $^{144}$Sm($^{40}$Ca,1p3n) and $\beta$-delayed fission fragments were measured with mica fission fragment detectors surrounding the rotating cylindrical target. ``...an examination of the data of [the table] in the light of the radioactive properties of the residual nuclei formed after particle emission from the compound system $^{184}$Pb with the initial excitation energy $E^* \sim$ (40$-$75) MeV leads to the assumption that fission with T$_{1/2}$ = 0.7 s occurs in the decay chain $^{180}$Tl $\to ^{180}$Hg$^*$.'' The half-life of 0.70$^{+0.12}_{-0.09}$~s is smaller than the currently accepted value of 1.5(2)~s, however, the observation of $\beta$-delayed fission was later confirmed with a half-life of 1.04(28)~s \cite{2010And01}.

\subsection*{$^{181}$Tl}
In the 1996 paper ``$\alpha$-decay properties of $^{181}$Pb'' Toth et al.\ described the identification of $^{181}$Tl \cite{1996Tot02}. A 400~MeV $^{92}$Mo beam from the Argonne tandem linac accelerator system bombarded a $^{90}$Zr target and $^{181}$Tl was formed in the fusion evaporation reaction $^{90}$Zr($^{92}$Mo,1p). Reactions were separated with the Argonne Fragment Mass Analyzer (FMA) and implanted in a double-sided silicon strip detector which also detected subsequent $\alpha$ decay. ``In [the figures], there are several close-lying groups slightly above 6.0~MeV. These are assigned to $^{177}$Au [6110 (65\%) and 6150 (35\%) keV] and to the 6180-keV transition associated with the new isotope $^{181}$Tl.'' A previous observation of an $\alpha$ decay in $^{181}$Tl was published by Bolshakov et al.\ only in a conference proceeding \cite{1993Bol01}. No half-life was reported, mentioning that the half-life was consistent with 3.4(6)~s, as reported by Bolshakov et al. Toth et al.\ reported a half-life of 3.2(3)~s two years later \cite{1998Tot01}. An earlier measurement of an $\alpha$-decay energy of 6566~keV and a half-life of 2.7~ms reported in a Ph.D. thesis was evidently incorrect \cite{1984Sch02}.

\subsection*{$^{182}$Tl}
The identification of $^{182}$Tl was reported in 1991 by Bouldjedri et al.\ in ``Identification and decay of $^{182}$Tl'' \cite{1991Bou01}. A thorium carbide target was bombarded with 600~MeV protons. $^{182}$Tl was produced in spallation reactions and identified with the CERN/ISOLDE facility. Charged particles and $\gamma$ rays were measured with a surface barrier Si and a Ge(Li) detector, respectively. Following the $\beta$ decay of $^{182}$Tl $\gamma$ transitions in $^{182}$Hg were measured: ``Four $\gamma$ transitions in $^{182}$Hg corresponding to the $8^+ \rightarrow 6^+ \rightarrow 4^+ \rightarrow 2^+ \rightarrow 0^+$ sequence have been observed. From the decay curves of the main $\gamma$ rays we deduced the weighted value for the half-life of T$_{1/2}$ = 3.1$\pm$1.0~s.'' This half-life corresponds to the presently adopted value. A previous tentative assignment of an $\alpha$ decay to $^{182}$Tl \cite{1986Kel01} was evidently incorrect.

\subsection*{$^{183}$Tl}
$^{183}$Tl was first observed in 1980 by Schrewe et al.\ in ``Alpha decay of neutron-deficient isotopes with 78 $\le$ Z $\le$ 83 including the new isotopes $^{183,184}$Pb and $^{188}$Bi'' \cite{1980Sch01}. A 6.25 MeV/nucleon $^{48}$Ti beam from the GSI UNILAC accelerator impinged on an enriched $^{142}$Nd target. $^{183}$Tl was formed in the fusion evaporation reaction $^{142}$Nd($^{48}$Ti,1p6n) and stopped in a FEBIAD ion source. Following reionization and mass separation, the ions were implanted into a carbon foil and their $\alpha$-decay was recorded. ``In addition to the known alpha line of $^{183}$Au and of $^{183}$Hg and its alpha decay daughter $^{179}$Pt, new high-energy alpha lines were observed. They were assigned to $^{183}$Tl confirming earlier unpublished data, and to the new isotope $^{183}$Pb.'' The measured half-life of 60(15)~ms corresponds to an isomeric state. The unpublished data mentioned in the quote referred to an unpublished thesis \cite{1978Mat01}. The ground state was first observed 19 years later \cite{1999Bat01}.

\subsection*{$^{184,185}$Tl}
In the 1976 paper ``Observation of $\alpha$-decay in thallium nuclei, including the new isotopes $^{184}$Tl and $^{185}$Tl'' by Toth et al.\ reported first evidence of $^{184}$Tl and $^{185}$Tl \cite{1976Tot01}. The Oak Ridge isochronous cyclotron accelerated $^{14}$N to 168 MeV which then impinged on WO$_3$ targets enriched in $^{180}$W. $^{184}$Tl and $^{185}$Tl were produced in (10n) and (9n) fusion-evaporation reactions, respectively, and identified in the UNISOR isotope separator facility. ``The 5.97 MeV $\alpha$-group seen at A = 185 is assigned to the new isotopes $^{185}$Tl. At A = 184 two $\alpha$-groups, 5.99 and 6.16 MeV, are assigned to the new isotope $^{184}$Tl because they decay with the same half-life.'' The measured half-life of 10(2)~s for $^{184}$Tl agrees with the presently accepted value of 9.7(6)~s, and the 1.7(2)~s half-life for $^{185}$Tl corresponds to an isomeric state. It is interesting to note that less than four weeks after the submission of the paper, seven of the co-authors were also co-authors on a submission reporting the ``New isotope $^{184}$Tl'' \cite{1976Col01} without referencing the first paper. The ground state of $^{185}$Tl has only been reported in conference proceedings \cite{1991Bol01,1993Bol01}.

\subsection*{$^{186}$Tl}
Hamilton et al.\ announced the discovery of $^{186}$Tl in the 1975 article ``Crossing of near-spherical and deformed bands in $^{186,188}$Hg and new isotopes $^{186,188}$Tl'' \cite{1975Ham01}. The Oak Ridge isochronous cyclotron accelerated $^{16}$O to 143$-$145 MeV to bombard a tantalum target to produce $^{186}$Tl in the (9n) fusion-evaporation reaction. Conversion electrons and $\gamma$-rays were measured with Si(Li) and Ge(Li) detectors. ``We have identified the new isotopes $^{186,188}$Tl with T$_{1/2}^{186}$=4.5$^{+1.0}_{-1.5}$ and 28$\pm$2~sec and T$_{1/2}^{188}$=71$\pm$1~sec, respectively.'' These values agree with the currently accepted values of 27.5(10)~s and 2.9(2)~s, for the ground state and and isomeric state, respectively. A year earlier Hamilton et al.\ had reported a 48(3)~s half-life in a first overview of the UNISOR program \cite{1974Ham01}.

\subsection*{$^{187}$Tl}
In the 1976 paper ``Observation of $\alpha$-Decay in thallium nuclei, including the new isotopes $^{184}$Tl and $^{185}$Tl'' by Toth et al.\ reported first evidence of $^{187}$Tl \cite{1976Tot01}. The Oak Ridge isochronous cyclotron accelerated $^{14}$N to 168 MeV which then impinged on $^{180}$W and $^{182}$W targets. $^{187}$Tl was produced in fusion-evaporation reactions and identified in the UNISOR isotope separator facility. ``Two isomers exist in $^{187}$Tl, with half-lives of about 18 and 40~s. We observed $\alpha$-decay only for the 18~s species.'' The half-life of 18(3)~s corresponds to an isomer. The reference to two known isomers in the quote refers to unpublished results. The observation of the ground state has only been reported in an conference abstract \cite{1980Woo01}.

\subsection*{$^{188}$Tl}
The first observation of $^{188}$Tl by Vandlik et al.\ was reported in the 1970 paper ``The new isotopes $^{190}$Tl and $^{188}$Tl'' \cite{1970Van01}. PbF$_2$ was bombarded with 660~MeV protons from the Dubna synchrocyclotron. The thallium fraction was chemically separated and $\gamma$ spectra were measured with a Ge(Li) detector. Results were given in a table, assigning a half-life of 1.6$^{+1}_{-0.5}$~min to $^{188}$Tl. This value agrees with the currently adopted value of 71(2)~s.

\subsection*{$^{189}$Tl}
Vandlik et al.\ reported the discovery of $^{189}$Tl in the 1972 paper ``The new isotope $^{189}$Tl'' \cite{1972Van01}. 660~MeV protons from the Dubna synchrocyclotron bombarded a PbF$_2$ target. Reaction products were separated with the online isotopes separation facility YaSNAPP and $\gamma$ rays were measured with a Ge(Li) detector. ``The agreement in the half-lives of the analyzed lines indicates, in all probability, that they belong to the decay of one and the same isotope having a state with a half-life T$_{1/2}$ = 1.4$\pm$0.4~min. In view of the procedure used to obtain this isotope, it is identified as Tl$^{189}$.'' This half-life corresponds to an isomeric state. The ground state was observed for the first time two years later \cite{1974Ham01}.

\subsection*{$^{190}$Tl}
The first observation of $^{190}$Tl by Vandlik et al.\ was reported in the 1970 paper ``The new isotopes $^{190}$Tl and $^{188}$Tl'' \cite{1970Van01}. PbF$_2$ was bombarded with 660~MeV protons from Dubna synchrocyclotron. The thallium fraction was chemically separated and $\gamma$ spectra were measured with a Ge(Li) detector. ``We conclude from the above results and the analogy with the adjacent isotopes that $^{190}$Tl has a metastable state as well as a ground state, whose half-lives are respectively 2.9$\pm$0.4~min and $\geqslant$3.6~min.'' These values agree with the currently accepted values of 2.6(3)~min and 3.7(3)~min for the ground state and an isomeric state, respectively. Vandlik et al.\ had the order of the states reversed.

\subsection*{$^{191}$Tl}
Vandlik et al.\ observed $^{191}$Tl and published their results in the 1974 paper ``Investigation of the decay $^{191}$Tl$\rightarrow^{191}$Hg$\rightarrow^{191}$Au'' \cite{1974Van01}. A lead target was bombarded with 660~MeV protons forming $^{191}$Tl in spallation reactions. Conversion electron- and $\gamma$-ray spectra were measured following chemical and mass separation. ``We cannot identify definitely the $^{191}$Tl level to which the observed 5.22-minute activity belongs. It is more probable, however, that it belongs to the ground state.'' This half-life is the currently adopted value. An earlier reported 10-min half-life \cite{1960Cha01} could not be confirmed and it was suspected that the data were contaminated by the 11.4~min half-life of $^{192}$Tl \cite{1961And01}.

\subsection*{$^{192}$Tl}
$^{192}$Tl was discovered in the 1961 paper ``Thallium isotopes 192 and 193'' \cite{1961And01}. Protons with energies between 75 and 105 MeV accelerated by the Uppsala synchro-cyclotron bombarded mercury targets. $^{192}$Tl was identified in an electro-magnetic mass separator following chemical separation. ``With a $^{192}$Tl sample prepared as described in [the section], the prominent L$_{II}$ line of a 109~keV transition and the K and L lines of a (424$\pm$1)~keV E2 transition (K/L $\sim$ 2.5) were found, all showing an about 12~min period.'' The reported half-life of 11.4(14)~min corresponds to an isomeric state. The ground state was first identified fourteen years later \cite{1975Van02}.

\subsection*{$^{193}$Tl}
In 1960 K.F. and G.A. Chackett reported the first observation of $^{193}$Tl in ``New light isotopes of thallium produced by bombardment of tungsten with nitrogen ions'' \cite{1960Cha01}. Enriched $^{184}$W targets were bombarded with 60~MeV $^{14}$N from the Nuffield cyclotron at Birmingham, England and $^{193}$Tl was formed in the (5n) fusion-evaporation reaction. Following chemical separation, the activity was measured with end-window halogen-quenched G.M. tubes. ``The bombardment of $^{184}$W gave a thallium fraction which decayed in 30 min, 4.0 hr and 17 hr periods, confirming that 30 min $^{193}$Tl is the parent of 4 hr $^{193}$Hg.'' The reported half-life of 30(3)~m is within a factor of two of the currently accepted value of 21.6(8)~m.

\subsection*{$^{194}$Tl}
In the 1960 paper ``Low mass odd-odd isomers of thallium'', Jung and Andersson described the discovery of $^{194}$Tl \cite{1960Jun01}. Hg$_2$Cl$_2$ targets were bombarded with 80$-$90~MeV protons from the Uppsala synchro-cyclotron. $^{194}$Tl was produced in (p,xn) reactions and identified by mass separation following chemical separation. ``Thus the only conclusion we can draw is that, within the limits of error, the ground state and the 7+ isomeric state of Tl$^{194}$ have the same half-life, (33.0$\pm$0.5) min and (32.8$\pm$0.2) min, respectively.'' These half-lives correspond to the presently accepted values.

\subsection*{$^{195}$Tl}
Knight and Baker discovered $^{195}$Tl in the 1955 paper ``Radiochemical study of Tl$^{195}$, Tl$^{197}$, and Tl$^{198m}$'' \cite{1955Kni01}. The Brookhaven 60-inch cyclotron was used to bombard enriched $^{196}$Hg targets with 20-MeV deuterons. The resulting decay curves were measured following chemical separation. ``The decay of the Tl$^{195}$ and Tl$^{197}$, as obtained from the activities of their mercury daughters, is plotted in [the figure]. The Tl$^{195}$ half-life was found to be 1.2$\pm$0.1 hours.'' This half-life agrees with the presently accepted value of 1.16(5)~h.

\subsection*{$^{196,197}$Tl}
The discovery of $^{196}$Tl and $^{197}$Tl was announced in 1955 by Andersson et al.\ in ``Neutron deficient isotopes of Pb and Tl-III:\ mass numbers below 200'' \cite{1955And01}. A thallium target was bombarded with protons from the Uppsala synchrocyclotron. Activities were measured in a two-directional focusing $\beta$-spectrometer. ``$^{197}$Tl.$-$The 134~kev $\gamma$-ray found in the decay of $^{197}$Tl is undoubtedly identical with the $\gamma$-ray of the same energy emitted in the decay of $^{197m}$Hg. $^{196}$Tl.$-$The assignment of the 426~kev $\gamma$-ray to $^{196}$Tl is supported by the following facts. Excitation relations indicated A$<$198. In the electron spectrum of A=197 (mass separated), however, the conversion lines of the 426~kev $\gamma$-ray did not show up, in spite of their strong intensities. Furthermore a 426~kev $\gamma$-ray has been observed in the $\beta$-decay of $^{197}$Au.'' The very approximate 4~h half-life for $^{196}$Tl was measured more accurately as 2.4(1)~h two years later \cite{1957And01}. This value and the half-life of 2.8$\pm$0.4~h for $^{197}$Tl are close to the currently accepted values of 1.84(3)~h and 2.84(4)~h, respectively.

\subsection*{$^{198-200}$Tl}
$^{198}$Tl, $^{199}$Tl and $^{200}$Tl were discovered by Orth et al.\ as reported in ``Radioactive thallium isotopes produced from gold'' \cite{1949Ort01}. The Berkeley 60-inch cyclotron was used to irradiate gold targets with 38, 28, and 20~MeV $^4$He beams to form $^{198}$Tl, $^{199}$Tl and $^{200}$Tl in ($\alpha$,3n), ($\alpha$,2n), and ($\alpha$,n), respectively. Decay curves were recorded following chemical separation. ``The maximum yields of 1.8-hour, 7-hour, and 27-hour thallium were obtained at 38 Mev, 28 Mev, and 20 Mev, respectively. The thresholds were in the same decreasing order but were not well defined by these experiments. These data suggest ($\alpha$,3n), ($\alpha$,2n), and ($\alpha$,n) reactions leading to Tl$^{198}$, Tl$^{199}$, and Tl$^{200}$, respectively.'' The half-life for $^{198}$Tl corresponds to an isomeric state and the half-lives of $^{199}$Tl and $^{200}$Tl agree with the currently adopted values of 7.42(8)~h and 26.1(1)~h, respectively. Previously reported half-lives of 10.5~h and 44~h assigned to $^{200}$Tl and/or $^{201}$Tl \cite{1940Kri01} were evidently incorrect.

\subsection*{$^{201}$Tl}
In 1950 Neumann and Perlman described the first observation of $^{201}$Tl in ``Isotopic assignments of bismuth isotopes produced with high energy particles'' \cite{1950Neu01}. Lead targets were bombarded with 100 MeV protons and deuterons from the Berkeley 184-inch cyclotron and bismuth isotopes were chemically separated. $^{201}$Tl was identified from the decay of $^{201}$Bi. ``When this experiment was done, a new decay chain consisting of an 8-hr.\ lead and 72-hr.\ thallium was indeed found. The thallium in particular was best assigned to Tl$^{201}$, because Tl$^{198}$, Tl$^{199}$, and Tl$^{200}$ have been assigned to other activities, while Tl$^{202}$ is not only assigned to a different activity, but would be blocked by long-lived Pb$^{202}$.'' The assigned half-life of 72(3)~h agrees with the presently accepted value of 3.0421(17)~d. Previously reported half-lives of 10.5~h and 44~h assigned to $^{200}$Tl and/or $^{201}$Tl \cite{1940Kri01} were evidently incorrect.

\subsection*{$^{202}$Tl}
First evidence of $^{202}$Tl was reported in 1940 by Krishnan and Nahum in ``Deuteron bombardment of the heavy elements I. Mercury, thallium and lead'' \cite{1940Kri01}. The Cavendish cyclotron was used to bombard mercuric oxide with 9 MeV deuterons. $^{202}$Tl was identified by chemical separation and the activities were measured with a Geiger counter and a scale-of-eight thyratron counter. ``A 13 day electron active body was obtained as a product of the deuteron bombardment of mercury. This was chemically identified as due to a thallium isotope. This isotope is also formed as a product of fast neutron bombardment of thallium, showing thereby that it should be assigned to Tl$^{202}$.'' This half-life agrees with the presently adopted value of 12.31(8)~d.

\subsection*{$^{203}$Tl}
Aston reported the observation of $^{203}$Tl in the 1931 paper ``Constitution of thallium and uranium'' \cite{1931Ast02}. The mass spectrum of $^{203}$Tl was measured with the Cavendish mass spectrometer. ``The lines of thallium were obtained by means of its triethyl compound. As was expected, it consists of two isotopes, 203 and 205. The latter predominates to an amount in good agreement with the chemical atomic weight (204.39).''

\subsection*{$^{204}$Tl}
Harbottle reported the observation of $^{204}$Tl in the 1953 paper ``The half-life of thallium-204'' \cite{1953Har02}. Neutrons irradiated thallous nitrate in the Oak Ridge Reactor and the products were cooled for 2.2~y. The source was then counted for 2.9~y in a methane-flow proportional counter. ``A least-squares fit of the results, plotted as log (Tl$^{204}$ counting rate corrected for background) versus time, yields a half-life of 4.02$\pm$0.12~years. This half-life is not far from the earlier value of 3.5~years but is entirely inconsistent with the later values of 2.7~years.'' This value agrees with the currently accepted half-life is 3.783(12)~y. The 3.5~y half-life mentioned in the quote had been assigned to $^{206}$Tl instead of $^{204}$Tl \cite{1941Faj01} and the 2.7~y half-life referred to a private communication listed in the 1953 Table of Isotopes \cite{1953Hol01}. Previously measured half-lives of 4.1~min \cite{1937Hey01} 4.4(1)~min \cite{1940Kri01} were evidently incorrect.

\subsection*{$^{205}$Tl}
Aston reported the observation of $^{205}$Tl in the 1931 paper ``Constitution of thallium and uranium'' \cite{1931Ast02}. The mass spectrum of $^{205}$Tl was measured with the Cavendish mass spectrometer. ``The lines of thallium were obtained by means of its triethyl compound. As was expected, it consists of two isotopes, 203 and 205. The latter predominates to an amount in good agreement with the chemical atomic weight (204.39).''

\subsection*{$^{206}$Tl}

In 1935 Preiswerk and von Halban reported the discovery of $^{206}$Tl in the paper ``Sur quelques radio\'el\'ements produits par les neutrons'' \cite{1935Pre01}. Metallic thallium and thallium nitrate were irradiated with neutrons from an 800~mCi radon-beryllium source. ``En plus de l'\'emission de p\'eriode 97 minutes, signal\'ee par J.C. Mac Lennan, L.G. Grimmet et J. Read \cite{1935McL01}, nous avons observ\'e une activit\'e de faible intensit\'e de periode 4 minutes. Les essais chimiques montrent que le corps activ de courte p\'eriode ne pr\'ecipite pas avec Au par H$^2$S en milieu acide, ni avec Hg et Pb dans les memes conditions. L'intensit\'e de l'\'emission de longue p\'eriode \'etait trop faible pur permettre une analyse chimique. Mais, comme d'autre part les quantit\'es form\'ees des deux corps sont augment'ees lorsque les neutrons sont ralentis par la paraffine, il est tr\`es vraisemblable que les corps produits sont isotopes du thallium, form\'es par capture d'un neutron, l'un \`a partir du $^{203}$Tl et l'autre `a partir du $^{205}$Tl: $^{203}_{81}$Tl + $^1_0$n = $^{204}_{81}$Tl; $^{204}_{81}$Tl$\xrightarrow[97 min.]{\beta} ^{204}_{82}$Pb $-$ $^{205}_{81}$Tl + $^1_0$n = $^{206}_{81}$Tl; $^{206}_{81}$Tl$\xrightarrow[4 min.]{\beta} ^{206}_{82}$Pb.'' [In addition to the emission of a 97 minutes period, reported by J.C. Mac Lennan, L.G. Grimmet et J. Read \cite{1935McL01}, we observed a low-intensity activity with a half-life of 4 minutes. Chemical tests show that the source of short half-life does not precipitate with gold in H$_2$S, nor with Hg and Pb under the same conditions. The intensity of the decay of the long half-life was too low to allow a pure chemical analysis. But like the other, the amount formed of the two activities are increased when the neutrons are slowed down by paraffin, it is very likely that the sources of these activites are isotopes of thallium, formed by neutron capture, one from $^{203}$Tl and the other one from $^{205}$Tl: $^{203}_{81}$Tl + $^1_0$n = $^{204}_{81}$Tl; $^{204}_{81}$Tl$\xrightarrow[97 min.]{\beta}^ {204}_{82}$Pb $-$ $^{205}_{81}$Tl + $^1_0$n = $^{206}_{81}$Tl; $^{206}_{81}$Tl$\xrightarrow[4 min.]{\beta}^ {206}_{82}$Pb.''] The measured half-life of 4~min assigned to $^{206}$Tl agrees with the presently adopted value of 4.202(11)~min.

\subsection*{$^{207}$Tl}
The first measurement of the activity due to $^{207}$Tl was reported in 1908 by Hahn and Meitner in the paper ``Aktinium C, ein neues kurzlebiges Produkt des Aktiniums'' \cite{1908Hah02}. An actinium sample was used to measure the $\alpha$- and $\beta$-ray activities of the decay products separately. Following the separation of the $\beta$ emitter $^{211}$Pb (AcA, later reclassified as AcB), a new $\beta$ emitter with the half-life of 5.10~min was observed in addition to the known 2.15~min $\alpha$ emitter $^{211}$Bi (AcB, later reclassified as AcC). ``Das neue Produkt Aktinium C wurde auf verschiedene Weise von Aktinium A getrennt hergestellt und seine Zerfallsperiode zu 5,10 Min. bestimmt. Es sendet die bis jetzt dem Aktinium B zugeschriebenen $\beta$(+$\gamma$)-Strahlen aus, emittiert aber keine $\alpha$-Strahlen.'' [The new actinium product, actinium C was separated from actinium A in several ways and its half-life was determined to be 5.10~min. It emits $\beta$(+$\gamma$) rays which had been assigned to actinium B, but does not emit $\alpha$ rays.] This actinium C was later reclassified as actinium C'' and corresponds to $^{207}$Tl which has a currently adopted half-life of 4.77(3)~min.

\subsection*{$^{208}$Tl}
In the 1909 article ``Eine neue Methode zur Herstellung radioaktiver Zerfallsprodukte; Thorium D, ein kurzlebiges Produkt des Thoriums,'' Hahn and Meitner announced the discovery of a new activity in the thorium decay chain later identified as $^{208}$Tl \cite{1909Hah02}. The active precipitate from 0.5~g radio-thorium was studied by collecting ThC recoils from the ThB $\alpha$ decay on an electrode. ``Die aufgenommenen Zerfallskurven ergeben in guter \"Ubereinstimmung den Wert von 3,1 Minute als Zerfallsperiode. Die angef\"uhrten Tatsachen beweisen: 1. da\ss\ in dem gefundenen $\beta$ Strahlenprodukt ein neues radioaktives Element vorliegt, 2. da\ss\ es in genetischer Beziehung zum ThA steht, und 3. da\ss\ es nach ThB+C kommen mu\ss , da sonst ein der Nachbildung von ThB+C entsprechender $\alpha$-Anstieg h\"atte beobachtet werden m\"ussen. Die neue Substanz mu\ss\ daher als Thorium D bezeichnet werden.'' [The measured decay curves consistently result in a 3.1~min half-life. These facts demonstrate (1) that a new radioactive element must be present in the $\beta$ emitter, (2) that it is in a genetic relationship to ThA, and (3) that it must occur after ThB+C, because otherwise the increase in $\alpha$ activity due to reforming of ThB+C should have been observed. The new substance must therefore be called thorium D.] ThD was later reclassified as ThC''. The measured half-life agrees with the presently adopted value 3.053(4)~min.

\subsection*{$^{209}$Tl}
$^{209}$Tl was identified in 1950 by Hagemann in ``Properties of Tl$^{209}$'' \cite{1950Hag01}. A solution of $^{225}$Ac was separated from a $^{233}$U source. Decay curves of $^{209}$Tl were then measured following a second fast chemical separation. ``The half-life as determined by least-squares analysis of seven such decay curves was 2.20 min., with a probable error of $\pm$0.07 min.'' This half-life corresponds to the currently accepted value. The existence of $^{209}$Tl as part of the (4n+1) radioactive decay series had been shown earlier by the $\alpha$-decay of $^{213}$Bi \cite{1947Hag01,1947Eng01}.

\subsection*{$^{210}$Tl}
Hahn and Meitner reported a new activity in the radium decay chain in the 1909 article ``Nachweis der komplexen Natur von Radium C'' \cite{1909Hah01}. Radium B and radium C activities were chemically separated and measured with a $\beta$-radiation electroscope. Decay curves and aluminum absorption curves were measured. The activities were also separated with the $\alpha$ recoil method. ``Es wurden nach der R\"ucksto\ss methode Aktivit\"aten hegestellt, deren Zerfallsperiode zwischen 1 und 2.5 Minuten schwankte. Radium C besteht also mindestens aus zwei Substanzen, Radium C$_1$ und Radium C$_2$.'' [Activities with half-lives varying between 1 and 2.5 minutes were created with the recoil method. Thus, radium C consist at least of two substances, radium C$_1$ and radium C$_2$.] The activity was later termed RaC'' and the measured half-life is consistent with the presently adopted value of 1.3(3)~min for $^{210}$Tl.

\subsection*{$^{211,212}$Tl}
Pf\"utzner et al.\ reported the discovery of $^{211}$Tl and $^{212}$Tl in the 1998 publication ``New isotopes and isomers produced by the fragmentation of $^{238}$U at 1000 MeV/nucleon'' \cite{1998Pfu01}. A 1000 MeV/nucleon $^{238}$U beam from the SIS facility at GSI bombarded a beryllium target and projectile fragments were identified with the fragment separator FRS in the standard achromatic mode. ``The nuclei $^{209}$Hg, $^{210}$Hg, $^{211}$Tl, $^{212}$Tl, $^{218}$Bi, $^{219}$Po and $^{220}$Po have been identified for the first time.''

\subsection*{$^{213}$Tl}
In the 2010 paper ``Discovery and investigation of heavy neutron-rich isotopes with time-resolved Schottky spectrometry in the element range from thallium to actinium'', Chen et al.\ described the discovery of $^{213}$Tl \cite{2010Che01}. A beryllium target was bombarded with a 670~MeV/u $^{238}$U beam from the GSI heavy-ion synchrotron SIS and projectile fragments were separated with the fragment separator FRS. The mass and half-life of $^{213}$Tl was measured with time-resolved Schottky Mass Spectrometry in the storage-cooler ring ESR. ``In [the figure] a typical frequency spectrum is shown including the new $^{213}$Tl isotope, known isotopes for which mass values have been measured for the first time in this experiment and those suitable for providing reference masses.'' The quoted half-life of 101$^{+484}_{-46}$~s is currently the only measured value for $^{213}$Tl.

\subsection*{$^{214-217}$Tl}
$^{214}$Tl, $^{215}$Tl, $^{216}$Tl, and $^{217}$Tl were discovered by Alvarez-Pol and the results were published in the 2010 paper ``Production of new neutron-rich isotopes of heavy elements in fragmentation reactions of $^{238}$U projectiles at 1A GeV'' \cite{2010Alv01}. A beryllium  target was bombarded with a 1~A GeV $^{238}$U beam from the GSI SIS synchrotron. The isotopes were separated and identified with the high-resolving-power magnetic spectrometer FRS. ``To search for new heavy neutron-rich nuclei, we tuned the FRS magnets for centering the nuclei $^{227}$At, $^{229}$At, $^{216}$Pb, $^{219}$Pb, and $^{210}$Au along its central trajectory. Combining the signals recorded in these settings of the FRS and using the analysis technique previously explained, we were able to identify 40 new neutron-rich nuclei with atomic numbers between Z=78 and Z=87; $^{205}$Pt, $^{207-210}$Au, $^{211-216}$Hg, $^{214-217}$Tl, $^{215-220}$Pb, $^{219-224}$Bi, $^{223-227}$Po, $^{225-229}$At, $^{230,231}$Rn, and $^{233}$Fr.''


\section{$^{179-220}$Pb}\vspace{0.0cm}

\begin{figure}
	\centering
	\includegraphics[scale=.7]{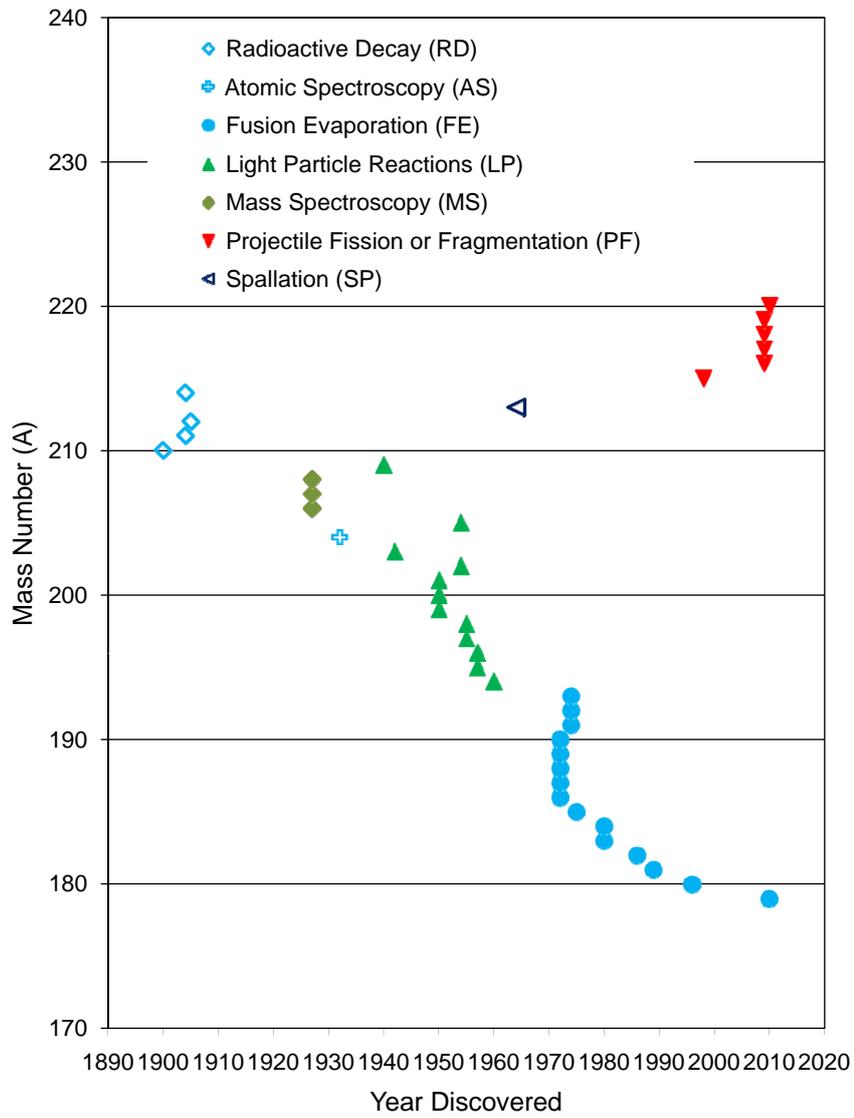}
	\caption{Lead isotopes as a function of time when they were discovered. The different production methods are indicated.}
\label{f:year-lead}
\end{figure}

Forty-two lead isotopes from A = 179--220 have been discovered so far; these include 4 stable, 26 neutron-deficient and 12 neutron-rich isotopes. According to the HFB-14 model \cite{2007Gor01} about 50 additional lead isotopes could exist. Figure \ref{f:year-lead} summarizes the year of first discovery for all lead isotopes identified by the method of discovery: radioactive decay (RD), mass spectroscopy (MS), atomic spectroscopy (AS), fusion evaporation reactions (FE), light-particle reactions (LP), projectile fission or fragmentation (PF), and spallation (SP). In the following, the discovery of each lead isotope is discussed in detail and a summary is presented in Table 1.

\subsection*{$^{179}$Pb}
Andreyev et al.\ announced the discovery of $^{179}$Pb in the 2010 paper ``The new isotope $^{179}$Pb and $\alpha$-decay properties of $^{179}$Tl$^m$'' \cite{2010And01}. A 232~MeV $^{40}$Ca beam from the GSI UNILAC bombarded enriched $^{144}$Sm targets forming $^{179}$Pb in the (5n) fusion-evaporation reaction. Residues were separated with the velocity filter SHIP and implanted in a position sensitive silicon strip detector, which also recorded subsequent $\alpha$ decay. ``The half-life value T$_{1/2}$($^{179}$Pb)=3.5$^{+1.4}_{-0.8}$~ms was deduced by using the data for the 12 correlated events.'' This is currently the only half-life measurement for $^{179}$Pb.

\subsection*{$^{180}$Pb}
$^{180}$Pb was first observed by Toth et al.\ in ``Identification of $^{180}$Pb'' in 1996 \cite{1996Tot01}. An enriched $^{144}$Sm target was bombarded with a 230-MeV $^{40}$Ca beam from the Berkeley 88-inch cyclotron and $^{180}$Pb produced in the fusion evaporation reaction $^{144}$Sm($^{40}$Ca,4n). The recoil products were deposited on a fast rotating catcher wheel and identified by their $\alpha$-decay. ``The $\alpha$ decay of the new isotope $^{180}$Pb was observed in $^{40}$Ca bombardments of $^{144}$Sm: E$_\alpha$ = 7.23(4)~MeV, and, T$_{1/2}$ = (4$^{+4}_{-2}$)~ms.'' This half-life agrees with the currently accepted value of 4.5(11)~ms. A previous attempt to produce $^{180}$Pb was unsuccessful \cite{1989Tot01}.

\subsection*{$^{181}$Pb}
Toth et al.\ identified $^{181}$Pb in the 1989 article ``Identification of $^{181}$Pb in $^{40}$Ca irradiations of $^{144}$Sm'' \cite{1989Tot01}. Enriched $^{144}$Sm targets were bombarded with 265 and 275~MeV $^{40}$Ca beams from the Berkeley 88-inch cyclotron producing $^{181}$Pb in the (3n) fusion-evaporation reaction. Reaction products ejected from the target were thermalized in helium gas and transported to a counting station on NaCl aerosols where $\alpha$ particles were measured with a Si(Au) surface barrier detector. ``Thus, with the observation of the 7044-keV $\alpha$ peak, which fits into the decay systematics for lead nuclei, we conclude that it is probably only the most intense of several $^{181}$Pb $\alpha$ transitions.'' The measured half-life of 50$^{+50}_{-30}$~ms agrees with the presently adopted value of 45(20)~ms. A previously reported $\alpha$ energy of 7211(10)~keV was evidently incorrect \cite{1986Kel01}.

\subsection*{$^{182}$Pb}
The first observation of $^{182}$Pb was reported in 1986 by Keller et al.\ in ``Cold fusion in symmetric $^{90}$Zr-induced reactions'' \cite{1986Kel01}. A $^{94}$Mo target was bombarded with a $^{90}$Zr beam from the GSI UNILAC facility forming $^{182}$Pb in the (2n) fusion evaporation reaction. Recoil products were identified with the velocity filer SHIP and implanted in two concentric surface-barrier detectors which also measured subsequent $\alpha$ decay. ``It was possible to identify some new $\alpha$-lines. The identification was based on cross-bombardments as well as on a comparison of excitation functions for several $\alpha$-lines...'' The measured $\alpha$-decay energy of 6921(10)~keV for $^{182}$Pb was later confirmed by Toth et al.\ (6919(15)~keV) \cite{1987Tot03} who, however, did not acknowledge the work by Keller et al.

\subsection*{$^{183}$Pb}
$^{183}$Pb was first observed in 1980 by Schrewe et al.\ in ``Alpha decay of neutron-deficient isotopes with 78 $\le$ Z $\le$ 83 including the new isotopes $^{183,184}$Pb and $^{188}$Bi'' \cite{1980Sch01}. A 6.25 MeV/nucleon $^{48}$Ti beam from the UNILAC accelerator at GSI impinged on an enriched $^{142}$Nd target. $^{183}$Pb was formed in the fusion evaporation reaction $^{142}$Nd($^{48}$Ti,7n) and stopped in a FEBIAD ion source. Following reionization and mass separation, the ions were implanted into a carbon foil and their $\alpha$-decay was recorded. ``In addition to the known alpha line of $^{183}$Au and of $^{183}$Hg and its alpha decay daughter $^{179}$Pt, new high-energy alpha lines were observed. They were assigned to $^{183}$Tl confirming earlier unpublished data, and to the new isotope $^{183}$Pb.''

\subsection*{$^{184}$Pb}
In the 1980 paper ``New isotope $^{184}$Pb'' Dufour et al.\ reported the discovery of $^{184}$Pb \cite{1980Duf01}. A 280~MeV $^{40}$Ca beam bombarded enriched samarium targets. Recoil products were stopped in a gas and guided onto a solid state detector with an electrostatic field. ``[The 6.62 MeV] $\alpha$ line can be assigned to the unknown isotope $^{184}$Pb, produced by the reactions ($^{40}$Ca,4n) in $^{148}$Sm and ($^{40}$Ca,3n) in $^{147}$Sm. A total number of 40 $\alpha$-decays have been observed, and the corresponding $\alpha$-energy is E$_\alpha$ = 6618$\pm$10~keV.'' The authors acknowledge a preprint of an independent observation of $^{184}$Pb which was submitted less than a month later \cite{1980Sch01}.

\subsection*{$^{185}$Pb}
Cabot et al.\ announced the first observation of $^{185}$Pb in the 1975 paper ``Ca induced reactions on $^{141}$Pr and $^{150}$Sm: new gold and lead isotopes $^{176}$Au, $^{175}$Au, $^{185}$Pb'' \cite{1975Cab01}. $^{150}$Sm targets were bombarded with 200$-$245~MeV $^{40}$Ca beams from the Orsay ALICE accelerator. Recoil nuclei were collected with a helium jet and subsequent $\alpha$ decay was measured. ``Thus we conclude that the 6.40 and 6.48~MeV activities are characteristic of the new isotope $^{185}$Pb produced by the (Ca,5n) reaction. This assignment is supported by the systematics of $\alpha$-decay for the Pb isotopes, as shown in [the figure].''

\subsection*{$^{186-190}$Pb}
The first observation of $^{186}$Pb, $^{187}$Pb, $^{188}$Pb, $^{189}$Pb, and $^{190}$Pb were described by Gauvin et al.\ in 1972 in ``$\alpha$ decay of neutron-deficient isotopes of bismuth and lead produced in (Ar,xn) and (Kr,xn) reactions'' \cite{1972Gau01}. The ALICE accelerator at Orsay was used to bombard a $^{155}$Gd target with 302$-$500~MeV $^{40}$Ar beams forming $^{186-190}$Pb in (9n-5n) fusion-evaporation reactions. Recoil products were identified with a helium jet technique and $\alpha$-decay spectroscopy. ``Two new lead isotopes were found: $^{187}$Pb, E$_\alpha$ = 6.08 MeV, t$_{1/2}$ = 17.5 sec; and $^{186}$Pb, E$_\alpha$ = 6.32 MeV, t$_{1/2}$ = 7.9 sec.'' The observation of $^{188}$Pb, $^{189}$Pb, and $^{190}$Pb was not considered a new discovery referring to an overview article by Eskola \cite{1967Esk01}, who listed results for these isotopes based on a private communication by Siivola. The measured half-lives of 7.9(16)~s for $^{186}$Pb, 17.5(36)~s for $^{187}$Pb, and 23.6(45)~s for $^{188}$Pb are in reasonable agreement with the presently accepted values of 4.82(3)~s, 15.2(3)~s, and 25.5(1)~s, respectively. For $^{189}$Pb and $^{190}$Pb only the $\alpha$ decay energies were measured.

\subsection*{$^{191,192}$Pb}
The first observation of $^{191}$Pb and $^{192}$Pb was reported by Le Beyec et al.\ in 1974 in ``New neutron deficient Pb and Bi nuclides produced in cross bombardments with heavy ions'' \cite{1974LeB01}. Fluorine and oxygen beams with energies up to 10.4 MeV/nucleon were accelerated by the Berkeley HILAC and bombarded self-supporting tantalum foils and WO$_3$ targets enriched in $^{182}$W, respectively. Excitation function were recorded and $\alpha$-decay energies and half-lives were measured. ``$^{191}$Pb: E$_{\alpha}$=5.29$\pm$0.02~MeV, (t$_{1/2}$=1.3$\pm$0.3~min) This $\alpha$ray was obtained in highest yield for an excitation energy of 120~MeV in the case of reactions induced in $^{181}$Ta by $^{19}$F ions. Also, in the $^{182}$W($^{16}$O,xn) series, the maximum occurred at an excitation energy of 90 MeV, corresponding to the emission of seven neutrons from the compound nucleus $^{198}$Pb. $^{192}$Pb: E$_{\alpha}$=5.06$\pm$0.03~MeV, (t$_{1/2}$=2.3$\pm$0.5~min) In the case of Bi, the excitation energy necessary for the emission of eight neutrons was around 100~MeV. Since the neutron binding energies for $^{192}$Pb to $^{200}$Pb are very close to binding energies for $^{193}$Bi$-^{201}$Bi, one might deduce that at E$^*$ = 105 MeV, the peak of the excitation function of the 5.06-MeV $\alpha$ ray obtained in the $^{181}$Ta($^{19}$F,xn)-induced reaction corresponds to x=8, and therefore to $^{192}$Pb.'' The measured half-lives of 1.3(3)~min ($^{191}$Pb) and 2.3(5)~min ($^{192}$Pb) are close to the currently accepted values of 1.33(8)~min and 3.5(1)~min, respectively. Similar results were already included in an overview article by Eskola \cite{1967Esk01}, quoting a private communication by Siivola.

\subsection*{$^{193}$Pb}
Newton et al.\ reported the first observation of $^{193}$Pb in the 1974 paper ``Rotational bands in the light odd-mass Tl nuclei'' \cite{1974New01}. A 124~MeV $^{19}$F beam from the Berkeley HILAC was incident on a $^{181}$Ta target forming $^{193}$Pn in the (7n) fusion-evaporation reaction. Conversion electrons and $\gamma$-ray spectra were measured. ``The only two lines which we could definitely assign to the decay of $^{193}$Pb were at 3392 and 366~keV... The half-lives for these two $\gamma$-rays were found to be 5.0$\pm$0.6~min and 5.8$\pm$0.3~min, respectively.'' These half-lives correspond to an isomeric state.

\subsection*{$^{194}$Pb}
In the 1960 paper ``Low mass odd-odd isomers of thallium'', Jung and Andersson described the discovery of $^{194}$Pb \cite{1960Jun01}. Protons of up to 90 MeV were accelerated by the Uppsala synchro-cyclotron and bombarded thallium targets. $^{194}$Pb was produced in (p,xn) reactions and identified by mass separation following chemical separation. ``In order to look for the ground state of Tl$^{194}$, a B sample was prepared. The earlier unknown Pb$^{194}$ turned out to have an 11 min half-life, measured on the K line of a 204 keV $\gamma$-ray converting in Tl and tentatively identified as the 2$- \to$ 2$-$ transition (K/L $>$ 4).'' The measured half-life of 11(2)~min agrees with the presently accepted value of 10.7(6)~min.

\subsection*{$^{195,196}$Pb}
Andersson et al.\ observed $^{195}$Pb and $^{196}$Pb as reported in the 1957 article ``Lead and thallium isotopes in the mass range 195-199'' \cite{1957And01}. Natural thallium targets were bombarded with 45$-$115~MeV protons from the Uppsala synchrocyclotron creating lead isotopes in (p,xn) reactions. Conversion electrons and $\gamma$-spectra were measured. ``The decay curve of a mass 195 sample could be decomposed into three components: 11.4~hr (probably identical with 9.5~hr Hg$^{195}$), 1.2~hr and $<$30~min. The 1.2~hr activity turned out to have a 37~keV $\gamma$-ray that is undoubtly identical with a transition earlier found in the decay of Hg$^{195m}$ and is thus assigned to Tl$^{195}$. Knight and Baker \cite{1955Kni01} report exactly the same half-life for this isotope, which they obtained by deuteron bombardment of Hg, enriched in mass 196. It was natural to expect the short-lived component to be associated with Pb$^{195}$... Decay curves of mass 196 samples showed a 40~min component, but this was not considered conclusive because contamination from mass 197 could not be entirely excluded. The assignment of a 37~min activity to Pb$^{196}$ is, however, supported by its relationship to Tl$^{196}$ as described below.'' The measured half-life of 17(1)~min measured for $^{195}$Pb corresponds to an isomeric state and the quoted half-life of 37(3)~min for $^{196}$Pb is the currently adopted value. The ground state of $^{195}$Pb was first measured 25 years later \cite{1982Hic01}.

\subsection*{$^{197,198}$Pb}
The discovery of $^{197}$Pb and $^{198}$Pb was announced in 1955 by Andersson et al.\ in ``Neutron deficient isotopes of Pb and Tl-III:\ mass numbers below 200'' \cite{1955And01}. A thallium target was bombarded by protons from the Uppsala synchrocyclotron. Conversion electrons were measured in a two-directional focusing $\beta$-spectrometer. ``$^{197}$Pb.$-$There are indications that the 169 and 234~keV $\gamma$-rays are converted in Pb, suggesting an isomeric state which may have about the same half-life as the ground state... $^{199}$Tl and and $^{198}$Tl.$-$The $\gamma$-rays found in the decay of these nuclides confirm earlier measurements. As to $^{198m}$Tl the $\gamma$-rays reported by Bergstrom et al.\ \cite{1953Ber01} and Passel et al.\ \cite{1954Pas01} were not observed. One possible explanation is that the isomeric state of $^{198}$Pb has not the half-life $\approx$25 minutes as suggested by Passel et al. but is so short-lived that it decays considerably before the chemical separation of Pb from the Tl target.'' The half-lives for $^{197}$Pb and $^{198}$Pb were listed in a table as 42(3)~m and 2.3(2)~h, respectively. The half-life for $^{197}$Pb corresponds to an isomeric state, while the half-life for $^{198}$Pb agrees with the currently accepted value of 2.4(1)~h. The ground state of $^{197}$Pb was first observed 24 years later \cite{1979Rap01}. Earlier assignments of a 25~min half-life to $^{198}$Pb \cite{1950Nau01,1951Kar01} were evidently incorrect.

\subsection*{$^{199-201}$Pb}
In 1950 Neumann and Perlman described the first observation of $^{199}$Pb, $^{200}$Pb, and $^{201}$Pb, in ``Isotopic assignments of bismuth isotopes produced with high energy particles'' \cite{1950Neu01}. Lead targets were bombarded with 100 MeV protons and deuterons from the Berkeley 184-inch cyclotron and $^{199}$Pb, $^{200}$Pb, and $^{201}$Pb were identified following chemical separation measuring activities with a mica end-window Geiger tube. ``In order to determine the half-life for Pb$^{199}$, one of the lead fractions was subjected to successive removals of thallium at 60-min. intervals, and the yields of the 7-hr. component were used to define the half-life of the Pb$^{199}$ parent. A value of $\sim$80 min. was obtained for the half-life of Pb$^{199}$ in this manner... The establishment of a 35-min.\ bismuth as the parent of the 18-hr.\ Pb$^{200}$ and 27-hr.\ Tl$^{200}$ was accomplished in a manner somewhat analogous to that already described in relation to [the figure] for the 12-hr.\ Bi$^{203}$-52-hr. Pb$^{203}$ pair... Finally, as shown in [the figure], the lead fraction removed from the bismuth fraction could be resolved into three components: 52-hr. Pb$^{203}$ in good yield, an 8-hr.\ period attributable to Pb$^{201}$, and a 68-min.\ period which is probably Pb$^{204m}$.'' These half-lives of $\sim$80~min ($^{199}$Pb), 18$\pm$3~h ($^{200}$Pb), and 8$\pm$2~h ($^{201}$Pb) agree with the presently accepted values of 90(10)~m, 21.5(4)~h, and 9.33(3)~h, respectively.

\subsection*{$^{202}$Pb}
The observation of $^{202}$Pb was reported by Maeder and Wapstra in the 1954 article ``A new isomer in lead'' \cite{1954Mae01}. Thallium was irradiated with 26~MeV deuterons and the resulting activity was studied with NaI scintillation spectrometers and a $\beta$-ray spectrometer following chemical separation. ``In irradiations of Tl with 26-Mev deuterons, a new 3.5$\pm$0.1~hr activity appeared in the Pb fraction. Its excitation curve, compared with those of 1.1-hr Pb$^{204\ast}$ and 2.3-day Pb$^{203}$ pointed to the reaction Tl$^{203}$(d,3n)Pb$^{202\ast}$.'' The given half-life corresponds to an isomeric state. The ground state was first observed eight months later \cite{1954Hui01}. An earlier assignment of a 5.6~s isomer to $^{202}$Pb \cite{1952Hop01} was evidently incorrect.

\subsection*{$^{203}$Pb}
Maurer and Ramm announced the discovery of $^{203}$Pb in the 1942 paper ``K\"unstlich radioaktive Isotope bei Blei und seinen Nachbarelementen unter Verwendung von Uran- und Thorblei'' \cite{1942Mau01}. Ordinary lead, as well as uranium- and thorium-lead were irradiated with fast neutrons produced from deuteron bombardment of lithium. Beta- and gamma-ray activities were measured following chemical separation. ``Nach diesem Ergebnis kann nur das Blei-Isotop der Masse 204 das Ausgangs-Isotop der Umwandlung sein, weil es im Vergleich zum gew\"ohnlichen Blei im Uran- und Thorblei nur in Spuren vorkommt. Die 52-Stunden-Periode ist also dem Proze\ss\ Pb$^{204}$(n,2n)Pb$^{203}$(52~Stunden) zuzuordnen.'' [Based on these results only the lead isotope of mass 204 can be the initial isotope for this transformation, because relative to ordinary lead it is present only in traces in uranium- and thorium-lead. The 52-hour period has thus to be assigned to the process Pb$^{204}$(n,2n)Pb$^{203}$(52~hours).] This half-life agrees with the presently accepted value of 51.92(3)~h. The earlier report of a small abundance of $^{209}$Pb by Aston \cite{1932Ast02} and the assignment of a 10.25~min half-life by Krishnan and Nahum \cite{1940Kri01} were evidently incorrect. Krishnan and Nahum assigned a 54~h half-life to $^{205}$Pb \cite{1940Kri01}.

\subsection*{$^{204}$Pb}
Sch\"uler and Jones reported first evidence of $^{204}$Pb in the 1932 paper ``\"Uber den spectroskopischen Nachweis einer neuen Blei-Isotope'' \cite{1932Sch01}. Spectroscopic studies of the hyperfine structure reviled the present of $^{204}$Pb. ``Unsere Beobachtungen an den Pb-I-Linien $\lambda\lambda$ 7228, 5201, und 5005 und der Pb-II-Linie $\lambda$ 5609 zeigen nun, da\ss\ in allen Strukturbildern eine \"uberz\"ahlige Komponente erscheint, die, im Hinblick auf den Isotopenverschiebungseffekt, immer dort liegt, wo die Isotope Pb$_{204}$ zu erwarten ist.'' [Our observations of the Pb-I-lines $\lambda\lambda$ 7228, 5201, and 5005 and the Pb-II-line $\lambda$ 5609 now show, that there is an additional component in all pictures, which occurs $-$ considering the isotope effect $-$ always at the location where one would expect the isotope Pb$^{204}$.] Aston submitted his mass spectroscopic results less than three months later \cite{1932Ast02}.

\subsection*{$^{205}$Pb}
Huizenga and Stevens reported their observation of $^{205}$Pb in the 1954 paper ``New long-lived isotopes of lead'' \cite{1954Hui01}. Lead isotopes were produced by bombardment of a thallium target with 21 MeV deuterons from the Argonne cyclotron. The irradiated sample was analyzed with a 12-inch 60$^\circ$ mass spectrometer following chemical separation. ``Since the Pb$^{205}$ yield in the second bombardment was probably comparable to the Pb$^{202}$ yield, the Pb$^{205}$ K- capture half-life is greater than 6$\times$10$^7$~years.'' The currently accepted half-life is 1.73$\times$10$^7$(7)~y. The earlier report of a small abundance of $^{205}$Pb \cite{1932Ast02} and the assignment of a 54~h \cite{1940Kri01} and 1.1~h half-life \cite{1950Gei01} were evidently incorrect.

\subsection*{$^{206-208}$Pb}
Aston discovered $^{206}$Pb, $^{207}$Pb, and $^{208}$Pb in 1927 in ``The constitution of ordinary lead'' \cite{1927Ast01}. A tetramethyl compound of lead was used in the Cavendish mass spectrometer. ``The vapour was first used diluted with carbon dioxide but later was admitted pure into the discharge tube. It works smoothly, but very long exposures are required. The three principal lines are 206 (4), 207 (3), 208 (7). The figures in brackets indicate roughly the relative intensities and are in good agreement with the atomic weight 207.2.''

\subsection*{$^{209}$Pb}
First evidence of $^{209}$Pb was described in 1940 by Krishnan and Nahum in ``Deuteron bombardment of the heavy elements I. Mercury, thallium and lead'' \cite{1940Kri01}. The Cavendish cyclotron was used to bombard mercuric oxide with 9 MeV deuterons. $^{209}$Pb was identified by chemical separation and the activities were measured with a Geiger counter and a scale-of-eight thyratron counter. ``The most intense activity observe in lead under deuteron bombardment has a half-life of 2.75$\pm$0.05 hr. It has been chemically identified as due to a lead isotope... This isotope is assigned to Pb$^{209}$, decaying to stable bismuth by an allowed $\beta$ disintegration.'' This half-life is consistent with the presently adopted value of 3.253(14)~h. A half-life of 3~h had previously been reported in a conference abstract \cite{1936Tho01}. An earlier report of a small abundance of $^{209}$Pb \cite{1932Ast02} was evidently incorrect.

\subsection*{$^{210}$Pb}
Radioactive lead was discovered in 1900 by Hofmann and Strauss and the results were published in the paper ``Radioactives Blei und radioactive seltene Erden'' \cite{1900Hof01}. Lead was chemically separated from various sources containing uranium. ``Wir fanden in verschiedenen Mineralien radioactives Blei und radioactive seltene Erden, die auch nach v\"olliger Trennung von Wismuth, resp.\ Thor und Uran ihre Wirksamkeit beibehielten... alle qualitativen und quantitativen Versuche ergaben, dass unsere activen Pr\"aparate nur Blei enthielten.'' [In various minerals we found radioactive lead and radioactive rare earth elements, which remained active even after complete separation from bismuth, thorium and uranium... every qualitative and quantitative experiment showed that the samples only contained lead.] Later it was recognized that this ``radiolead'' was the same as RaD ($^{210}$Pb) \cite{1904Rut02,1905Rut02,1905Rut03}.

\subsection*{$^{211}$Pb}
Debierne published the observation of a new activity in the actinium chain in the 1904 paper titled ``Sur l'\'emanation de l'actinium'' \cite{1904Deb01}. A new activity was observed following the decay of actinium emanation ($^{219}$Rn). ``J'ai \'egalement d\'etermin\'e la loi de d\'ecroissance de la radioacivit\'e induite provoqu\'ee par l'\'emanation de l'actinium en mesurant dans le m\^eme appareil l'activit\'e induite des condensateurs, depuis le moment o\`u l'\'emanation a cess\'e d'agir: La d\'ecroissance est r\'eguli\`ere, elle est de la moiti\'e en 40 minutes.'' [I also found the radioactive decay law caused by the actinium emanation by measuring the induced activity in the same setup, from the moment the emanation has ceased to act: The decay is regular, it reaches its half value in 40 minutes.] Rutherford reproduced the observation by Debierne and named it actinium A \cite{1905Rut01}. It was later reclassified as AcB. The measured half-life agrees with the currently accepted value of 36.1(2)~min.

\subsection*{$^{212}$Pb}
Rutherford reported an activity later identified as $^{212}$Pb in the 1905 paper ``Bakerian lecture.-The succession of changes in radioactive bodies'' \cite{1905Rut01}. The decay curves of ``excited activities'' following the decay of thorium emanation ($^{220}$Rn) were measured. ``The evidence, as a whole, thus supports the conclusion that the active deposit from thorium undergoes two successive transformations as follows: (1) A `rayless' change for which $\lambda_1$ = 1.75$\times$10$^{-5}$, i.e., in which half the matter is transformed in 11 hours; (2) A second change giving rise to $\alpha$, $\beta$ and $\gamma$ rays, for which $\lambda_2$ = 2.08$\times$10$^{-4}$, i.e., in which half the matter is transformed in 55 minutes.'' The measured half-life for the first decay which was named thorium A ($^{212}$Pb) agrees with the presently adopted value of 10.64(1)~h. Rutherford and Soddy had reported the observation of a ``Thorium-excited activity I'' from the decay of thorium emanation earlier \cite{1903Rut01}. ThA was later reclassified as ThB.

\subsection*{$^{213}$Pb}
The discovery of $^{213}$Pb was reported in 1964 by Butement et al.\ in ``A new isotope of lead: $^{213}$Pb'' \cite{1964But02}. A thorium target was bombarded with 370~MeV protons from the Liverpool cyclotron and $^{221}$Rn was produced in spallation reactions. The subsequent decay products were chemically separated and the resulting activity measured. ``The half life resolved from the growth in this experiment was 9.7 min, and its genetic relationship with the 48 min decay shows that the 9.7 min half-life must be that of $^{213}$Pb. The mean of three experiments gave a value of 10.2$\pm$0.3 min for the half-life of $^{213}$Pb.'' This half-life is the presently accepted value.

\subsection*{$^{214}$Pb}
In the 1904 paper ``Heating effect of the radium emanation'', Rutherford and Barnes described the discovery of a new activity later identified as $^{214}$Pb \cite{1904Rut01}. The decay curves of ``excited activities'' following the decay of radium and radium emanation ($^{222}$Rn) were measured. ``An analysis of the decay curves of excited activity, produced for different intervals of exposure in the presence of the emanation, shows that there are three well-marked changes occurring in emanation X of radium. In the first change, half the matter is transformed in 3 minutes; in the second, half in 34 minutes; and in the third, half in 28 minutes. A full account of the analysis of these changes and their peculiarities will be given by one of us in a later paper. The first change is accompanied only by $\alpha$ rays, the second change is not accompanied by $\alpha$ rays at all, and the third change by $\alpha$, $\beta$, and $\gamma$ rays.'' The 34~min half-life of the second decay - later named radium B - is close to the currently adopted half-life of 26.8(9)~min. Rutherford and Soddy had reported the observation of a ``Radium-excited activity II'' from the decay of thorium emanation earlier \cite{1903Rut01}.

\subsection*{$^{215}$Pb}
Pf\"utzner et al.\ observed $^{215}$Pb as described in the 1998 publication ``New isotopes and isomers produced by the fragmentation of $^{238}$U at 1000 MeV/nucleon'' \cite{1998Pfu01}. A 1~GeV/nucleon $^{238}$U beam from the SIS facility at GSI bombarded a beryllium target and projectile fragments were identified with the fragment separator FRS in the standard achromatic mode. The observation of $^{215}$Pb was not considered a discovery quoting a paper by Van Duppen et al.:``A complementary technique is the combination of high-energy proton-induced spallation of thick heavy targets with on-line isotope separation (ISOL), with which the isotopes $^{215}$Pb and $^{217}$Bi were recently observed \cite{1998Van01}.'' However, Van Duppen et al.\ did not actually observe $^{215}$Pb stating: ``In conclusion, we have presented a new method that allows detailed decay-spectroscopy studies of the neutron-rich `east of $^{208}$Pb' using the pulsed release from the ISOLDE targets... It has been successfully applied in a recent experiment where two new isotopes ($^{215}$Pb and $^{217}$Bi) were identified ...'' and referring to ``K. Rykaczewski {\it et al.}, to be published.'' This clearly does not constitute the discovery of $^{215}$Pb, thus the credit is given to Pf\"utzner et al.\ because $^{215}$Pb is cleanly identified in the mass-to-charge spectra.

\subsection*{$^{216-219}$Pb}
In 2009 Alvarez-Pol et al.\ reported the observation of $^{216}$Pb, $^{217}$Pb, $^{218}$Pb, and $^{219}$Pb in the paper ``Production cross-sections of neutron-rich Pb and Bi isotopes in the fragmentation of $^{238}$U'' \cite{2009Alv01}. The GSI SchwerIonen Synchrotron SIS was used to bombard a beryllium target with a 1 A GeV $^{238}$U beam. Reaction products were identified by mass and atomic number based on magnetic rigidity, time of flight, and energy loss. ``The high-resolving power FRagment Separator (FRS) at GSI allowed us to identify and determine the production cross-sections of 43 nuclei, nine of them for the first time ($^{216,217,218,219}$Pb and $^{219,220,221,222,223}$Bi).''

\subsection*{$^{220}$Pb}
$^{220}$Pb was discovered by Alvarez-Pol and the results were published in the 2010 paper ``Production of new neutron-rich isotopes of heavy elements in fragmentation reactions of $^{238}$U projectiles at 1A GeV'' \cite{2010Alv01}. A beryllium target was bombarded with a 1~A GeV $^{238}$U beam from the GSI SIS synchrotron. The isotopes were separated and identified with the high-resolving-power magnetic spectrometer FRS. ``To search for new heavy neutron-rich nuclei, we tuned the FRS magnets for centering the nuclei $^{227}$At, $^{229}$At, $^{216}$Pb, $^{219}$Pb, and $^{210}$Au along its central trajectory. Combining the signals recorded in these settings of the FRS and using the analysis technique previously explained, we were able to identify 40 new neutron-rich nuclei with atomic numbers between Z=78 and Z=87; $^{205}$Pt, $^{207-210}$Au, $^{211-216}$Hg, $^{214-217}$Tl, $^{215-220}$Pb, $^{219-224}$Bi, $^{223-227}$Po, $^{225-229}$At, $^{230,231}$Rn, and $^{233}$Fr.'' It is interesting to note that although the paper claims the discovery of additional lead isotopes, $^{215-220}$Pb, the observation of $^{216-219}$Pb had been reported a year earlier by the same authors \cite{2009Alv01} and the authors apparently also were aware of the previous observation of $^{215}$Pb by Pf\"utzner et al.\ as acknowledged in the 2009 paper by Alvarez-Pol et al. \cite{2009Alv01}.


\section{$^{184-224}$Bi}\vspace{0.0cm}

\begin{figure}
	\centering
	\includegraphics[scale=.7]{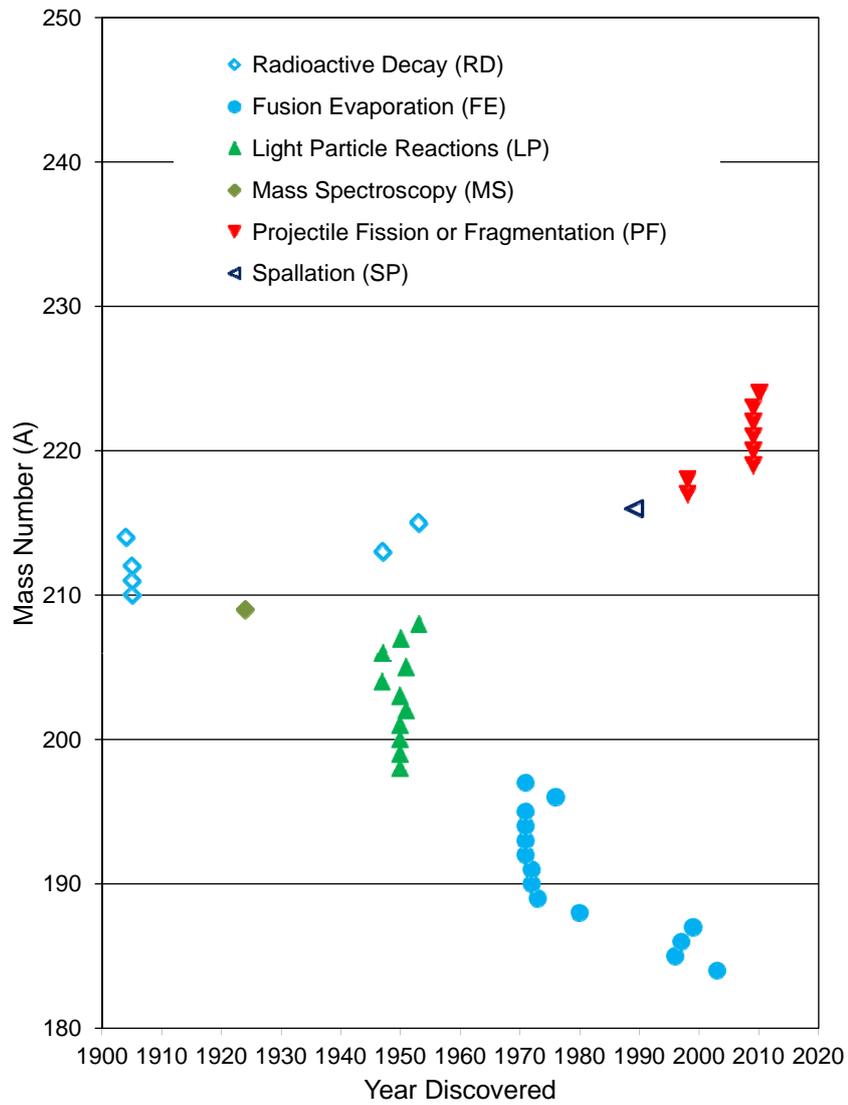}
	\caption{Bismuth isotopes as a function of time when they were discovered. The different production methods are indicated.}
\label{f:year-bismuth}
\end{figure}

Forty-one bismuth isotopes from A = 184--224 have been discovered so far; these include 1 stable, 25 neutron-deficient and 15 neutron-rich isotopes. According to the HFB-14 model \cite{2007Gor01} about 42 additional bismuth isotopes could exist. Figure \ref{f:year-bismuth} summarizes the year of first discovery for all bismuth isotopes identified by the method of discovery: radioactive decay (RD), mass spectroscopy (MS), fusion evaporation reactions (FE), light-particle reactions (LP), projectile fission or fragmentation (PF), and spallation (SP). In the following, the discovery of each bismuth isotope is discussed in detail and a summary is presented in Table 1.

\subsection*{$^{184}$Bi}
$^{184}$Bi was first observed in 2003 by Andreyev et al.\ in ``$\alpha$-decay spectroscopy of light odd-odd Bi isotopes - II: $^{186}$Bi and the new nuclide $^{184}$Bi'' \cite{2003And01}. A $^{94}$Mo beam was accelerated to 434$-$461~MeV by the GSI UNILAC heavy-ion accelerator and bombarded a $^{93}$Nb target forming $^{184}$Bi in the (3n) fusion-evaporation reaction. $^{184}$Bi was separated with the velocity filter SHIP and implanted in a 16-strip position-sensitive silicon detector which also measured subsequent $\alpha$ decay. ``Two $\alpha$-decaying isomeric states in $^{184}$Bi with half-life values of 12(2) ms and 6.6(1.5) ms were identified.'' These half-lives are presently the only measurements for $^{184}$Bi.

\subsection*{$^{185}$Bi}
Davids et al.\ discovered $^{185}$Bi in 1996 in ``Proton decay of an intruder state in $^{185}$Bi'' \cite{1996Dav01}. The Argonne ATLAS accelerator was used to bombard an enriched $^{95}$Mo target with 410~MeV $^{92}$Mo, and $^{185}$Bi was produced in the (1p1n) fusion-evaporation reaction. The isotopes were separated with the FMA fragment mass analyzer and implanted in a double-sided silicon strip detector which also measured subsequent proton emission. ``...the peak is assigned to the proton decay of $^{185}$Bi, with a corresponding Q value of 1.594 $\pm$ 0.009 MeV, and an associated cross section of $\sim$100~nb using an efficiency for the FMA of 20\%.'' The measured half-life of 44(16)~$\mu$s agrees with the presently accepted value of 49(7)~$\mu$s.

\subsection*{$^{186}$Bi}
In 1997 Batchelder et al.\ reported the first observation of ``The $\alpha$-decay properties of $^{186}$Bi'' \cite{1997Bat01}. A 420~MeV $^{92}$Mo beam from the Argonne ATLAS facility bombarded an enriched $^{97}$Mo target producing $^{186}$Bi in (p2n) fusion-evaporation reactions $^{186}$Bi was identified with the FMA fragment mass analyzer and implanted in a double-sided silicon strip detector which also measured subsequent $\alpha$ emission. ``In contrast to an earlier work wherein only one $\alpha$ transition was reported, we observed two transitions arising from two isomers in $^{186}$Bi with energy and half-lives of: 7158(20)~keV, T$_{1/2}$ = 15.0(17)~ms; and 7261(20)~keV, T$_{1/2}$ = 9.8(13)~ms.'' These half-lives are currently the only measurements. The reference to an earlier work corresponds to an unpublished thesis \cite{1984Sch02}.

\subsection*{$^{187}$Bi}
In the 1999 article ``Behavior of intruder based states in light Bi and Tl isotopes: The study of $^{187}$Bi $\alpha$ decay,'' Batchelder et al.\ announced the discovery of $^{187}$Bi \cite{1999Bat01}. $^{92}$Mo accelerated to 420~MeV at the Argonne ATLAS facility bombarded an enriched $^{97}$Mo target producing $^{187}$Bi in the (1p1n) fusion evaporation reaction. Recoils were separated with a fragment mass analyzer and implanted in a double-sided silicon strip detector which also recorded subsequent $\alpha$ decay. ``The previously unobserved $^{187}$Bi ground state (h$_{9/2}$) to $^{183}$Tl ground state (s$_{1/2}$) $\alpha$ transition was identified, establishing the $^{187}$Bi intruder state excitation energy to be 112(21)~keV, 70~keV less than that of the same level in $^{189}$Bi.'' The measured half-life of 32(3)~ms is included in the calculation of the currently accepted value. Earlier a half-life of 35(4)~ms half-life was only reported in an unpublished thesis \cite{1984Sch02}.

\subsection*{$^{188}$Bi}
$^{188}$Bi was first observed in 1980 by Schrewe et al.\ in ``Alpha decay of neutron-deficient isotopes with 78 $\le$ Z $\le$ 83 including the new isotopes $^{183,184}$Pb and $^{188}$Bi'' \cite{1980Sch01}. A 4.60 MeV/nucleon $^{84}$Kr beam from the UNILAC accelerator at GSI impinged on an enriched $^{107}$Ag target. $^{188}$Bi was formed in the fusion evaporation reaction $^{107}$Ag($^{84}$Kr,3n) and stopped in a FEBIAD ion source. Following reionization and mass separation, the ions were implanted into a carbon foil and their $\alpha$-decay was recorded. ``...alpha lines of 6632 keV and 6820 keV from reactions of $^{48}$Ti on $^{142}$Nd and $^{84}$Kr on $^{107}$Ag were assigned to the new isotopes $^{184}$Pb and $^{188}$Bi, respectively.'' The currently accepted half-life is 210(9)~ms.

\subsection*{$^{189}$Bi}
Gauvin et al.\ described the identification of $^{189}$Bi in the 1973 paper ``Reactions of $^{40}$Ar with $^{159}$Tb, $^{142}$Nd and $^{144}$Sm; new $\alpha$-activities from $^{189}$Bi, $^{173}$Pt and $^{177}$Au'' \cite{1973Gau01}. A terbium oxide target was bombarded with a 283~MeV $^{40}$Ar beam from the ORSAY ALICE accelerator to form $^{189}$Bi in the $^{159}$Tb($^{40}$Ar,10n) reaction. Recoil nuclei were collected by the helium gas-jet technique and subsequent $\alpha$-decay spectra were measured. ``Based on the fact that all of the observed counts in the 6.67~MeV peak occurred during the first counting interval of 2~sec, begun 1.5~sec after the end of the irradiation, we can set a conservative upper limit of $<$ 1.5~sec on the half-life of $^{189}$Bi.'' The currently adopted half-life is 658(47)~ms.

\subsection*{$^{190-191}$Bi}
The first observation of $^{190}$Bi and $^{191}$Bi were described by Gauvin et al.\ in 1972 in ``$\alpha$ decay of neutron-deficient isotopes of bismuth and lead produced in (Ar,xn) and (Kr,xn) reactions'' \cite{1972Gau01}. The ALICE accelerator at Orsay was used to bombard a $^{159}$Tb target with 302$-$500~MeV $^{40}$Ar beams forming $^{190}$Bi and $^{191}$Bi in (9n) and (8n) fusion-evaporation reactions, respectively. Recoil products were identified with a helium jet technique and $\alpha$-decay spectroscopy. ``We observed $\alpha$ emission from bismuth nuclides and isomers with A = 190$-$197 and from lead isotopes with A = 186$-$190.'' These observations were not considered new discoveries referring to an overview article by Eskola \cite{1967Esk01}, who listed results for these isotopes based on a private communication by Siivola. The measured half-lives of 5.4(5)~s for $^{190}$Bi and 12.0(7)~s for $^{191}$Bi are in agreement with the presently accepted values of 6.3(1)~s and 12.3(3)~s, respectively.

\subsection*{$^{192-195}$Bi}
$^{192}$Bi, $^{193}$Bi, $^{194}$Bi, and $^{195}$Bi were identified by Tarantin et al.\ in the 1970 paper ``Identification and study of the radioactive properties of bismuth isotopes with an electromagnetic mass separator in a heavy-ion beam'' \cite{1971Tar01}. A 200~MeV $^{20}$Ne beam from the Dubna U-300 cyclotron bombarded a $^{181}$Ta target forming $^{192-195}$Bi in (9n-6n) fusion-evaporation reactions. Recoil products were separated with an online mass separator and the subsequent $\alpha$ decay was measured with a semiconductor $\alpha$ counter. The $\alpha$-decay energies and half-lives are summarized in a table. For $^{192}$Bi only the decay energy (6.09(2)~MeV) was given. The half-lives of 62~s for $^{193}$Bi, 62~s for $^{194}$Bi, and 4~min for $^{195}$Bi, are consistent with the presently adopted values of 67(3)~s, 95(3)~s, and 183(4)~s, respectively. Earlier Treytl and Vali had assigned a 6.050(5)~MeV, 38(8)~s activity to either $^{191}$Bi or $^{195}$Bi, a 5.892(5)~MeV, 74(5)~s activity to either $^{192}$Bi or $^{196}$Bi and a 6.100(5)~MeV, 55(10)~s activity to either $^{191}$Bi or $^{195}$Bi. Tarantin et al.\ did not consider these observations new discoveries referring to an overview article by Eskola \cite{1967Esk01}, who listed results for these isotopes based on a private communication by Siivola.

\subsection*{$^{196}$Bi}
In the 1976 paper ``Levels in $^{194}$Pb and $^{196}$Pb populated in the decay of $^{194}$Bi and $^{196}$Bi'' Chojnacki et al.\ identified $^{196}$Bi \cite{1976Cho01}. $^{181}$Ta foils were bombarded with 110~MeV $^{20}$Ne and 150~MeV $^{22}$Ne beams from the Dubna heavy-ion cyclotron forming $^{196}$Bi in (7n) and (9n) fusion-evaporation reactions, respectively. Gamma-ray spectra and $\gamma$-$\gamma$ coincidences were measured with a Ge(Li) and a NaI(Tl) detector. ``The half-life of $^{196}$Bi as measured by us is T$_{1/2}$=4.6$\pm$0.5~min.'' This half-life agrees with the currently adopted value of 308(12)~s.

\subsection*{$^{197}$Bi}
$^{197}$Bi was identified by Tarantin et al.\ in the 1970 paper ``Identification and study of the radioactive properties of bismuth isotopes with an electromagnetic mass separator in a heavy-ion beam'' \cite{1971Tar01}. A 200~MeV $^{20}$Ne beam from the Dubna U-300 cyclotron bombarded a $^{181}$Ta target forming $^{197}$Bi in (4n) fusion-evaporation reactions. Recoil products were separated with an online mass separator and the subsequent $\alpha$ decay was measured with a semiconductor $\alpha$ counter. The $\alpha$-decay energies and half-lives are summarized in a table. The measured half-life of 9.5~min of a 5.81(2)~MeV $\alpha$ decay agrees with the currently accepted value of 9.3(5)~min. Tarantin et al.\ did not consider this observation a new discovery referring to an overview article by Eskola \cite{1967Esk01}, who listed results for these isotopes based on a private communication by Siivola.

\subsection*{$^{198-201}$Bi}
In 1950 Neumann and Perlman described the first observation of $^{198}$Bi, $^{199}$Bi, $^{200}$Bi, and $^{201}$Bi, in ``Isotopic assignments of bismuth isotopes produced with high energy particles'' \cite{1950Neu01}. Lead targets were bombarded with 100 MeV protons and deuterons from the Berkeley 184-inch cyclotron and $^{198}$Bi, $^{199}$Bi, $^{200}$Bi, and $^{201}$Bi were identified following chemical separation measuring $\alpha$- and $\beta$-activities with a mica end-window Geiger tube and a parallel plate chamber, respectively. ``7-Min.\ Bi$^{198(?)}$: ...When the short-lived thallium was resolved from the complex decay curve with 1.8-hr.\ half-life, the yields of this component indicated a half-life for the bismuth ancestor of 7~min... 25-Min.\ Bi$^{199}$: A bismuth of this period was first noted through its alpha-emission, and has now been assigned to Bi$^{199}$ through its genetic relationship to 7-hr. Tl$^{199}$ by way of electron capture decay... 35-Min.\ Bi$^{200}$: The assignment of this isotope is based upon the assignment of the 27-hr thallium to Tl$^{200}$. The decay sequence starts with 35-min. Bi$^{200}$, the daughter of which is an 18-hr. lead activity... Bi$^{201}$ Isomers: ...the half-life for the Bi$^{201}$ parent in the first experiments turned out to be about 90~min.\ rather than 62~min., the half-life of the alpha-activity thought to be Bi$^{201}$. The most promising solution at present is to assume that there are independently decaying isomers, one of about 1-hr half-life which has measurable alpha-branching, and another of about 2-hr.\ half-life which only exhibits electron capture decay.'' The measured half-lives of 7(1)~min ($^{198}$Bi), 25(5)~min ($^{199}$Bi), 35(5)~min ($^{200}$Bi), and 110(10)~min ($^{201}$Bi) agree with the presently accepted values of 10.3(3)~min, 27(1)~min, 36.4(5)~min, and 103(3)~min, respectively. The half-life of $^{201}$Bi is included in the calculation of the currently adopted value. Previously, 9~min, 27~min and 1$-$2~hr.\ half-lives was reported without a mass assignment \cite{1948Tem01}.

\subsection*{$^{202}$Bi}
$^{202}$Bi was discovered in ``Polonium isotopes produced with high energy particles'' by Karraker and Templeton in 1951 \cite{1951Kar01}. Natural lead and bismuth targets were bombarded with helium beams and protons, respectively, from the 184-in Berkeley cyclotron. Decay curves were measured following chemical separation. ``52-min Po$^{202}$ and 95-min Bi$^{202}$: In addition to the other bismuth activities mentioned above, we observed a 95-minute activity among the bismuth daughters of polonium produced in bombardments at fairly high energy.'' This half-life for $^{202}$Bi agrees with the currently accepted value of 1.71(4)~h.

\subsection*{$^{203}$Bi}
In 1950 Neumann and Perlman described the first observation of $^{203}$Bi, in ``Isotopic assignments of bismuth isotopes produced with high energy particles'' \cite{1950Neu01}. Lead targets were bombarded with 100 MeV protons and deuterons from the Berkeley 184-inch cyclotron. $^{203}$Bi was identified by decay curve measurements and correlations with the thallium daughter decay following chemical separation. ``12-hr. $^{203}$Bi: No attempt was made to resolve this period directly out of the bismuth fraction nor to determine radiation characteristics, since it has the same half-life as Bi$^{204}$ which was always present. However, the half-life was readily discerned by periodically removing the lead fraction; and upon resolution of 52-hr.\ Pb$^{203}$, it was found that its yield decreased with a half-life of 12$\pm$1~hr.'' This half-life for $^{203}$Bi agrees with the presently accepted value of 11.76(5)~h.

\subsection*{$^{204}$Bi}
Howland et al.\ observed $^{204}$Bi for the first time in 1947 in ``Artificial radioactive isotopes of polonium, bismuth and lead'' \cite{1947How01}. The Berkeley 60-inch cyclotron was used to bombard an enriched $^{204}$Pb target with a 40~MeV $^4$He beam and a thallium target with 20~MeV deuterons forming $^{204}$Bi in the (d,2n) and ($\alpha$,3n) reactions, respectively. Electrons and $\gamma$ rays were measured. ``The high yield of the 12-hour bismuth from deuterons on Pb$^{204}$ limits the assignment to Bi$^{203}$ or Bi$^{204}$, and the high yield from helium ions on thallium sets 204 as the lowest possible mass. Therefore the isotope is Bi$^{204}$.'' The measured half-life of 12~h agrees with the presently accepted value of 11.22(10)~h.

\subsection*{$^{205}$Bi}
$^{205}$Bi was discovered in ``Polonium isotopes produced with high energy particles'' by Karraker and Templeton in 1951 \cite{1951Kar01}. Natural lead and bismuth targets were bombarded with helium beams and protons and deuterons, respectively, from the 184-in Berkeley cyclotron. Conversion electrons and $\gamma$-rays were measured with a beta-spectrometer and a Geiger counter, respectively, following chemical separation. ``The half-life of Bi$^{205}$ produced by the decay of Po$^{205}$ was found to be 14.5 days.'' This half-life agrees with the currently accepted value of 15.31(4)~d.

\subsection*{$^{206}$Bi}
Howland et al.\ observed $^{206}$Bi for the first time in 1947 in ``Artificial radioactive isotopes of polonium, bismuth and lead'' \cite{1947How01}. The Berkeley 60-inch cyclotron was used to bombard lead targets with a 40~MeV $^4$He beam and thallium targets with 20~MeV deuterons. Electrons and $\gamma$ rays were measured. ``The 9-day Po$^{206}$ was found to decay into the 6.4-day bismuth activity assigned to Bi$^{206}$ or Bi$^{207}$ by Fajans and Voigt \cite{1941Faj01,1941Faj02}. The assignment to Bi$^{206}$ is in agreement with the observation of Corson, MacKenzie, and Segre that this activity is not produced by the alpha-decay of $_{85}$At$^{211}$ \cite{1940Cor01}.'' The measured half-life of 6.4~d for $^{206}$Bi agrees with the presently adopted value of 6.243(3)~d. In addition to the different earlier assignment mentioned in the quote, Krishnan had assigned a 6.35(20)~d half-life incorrectly to $^{208}$Bi \cite{1940Kri01}.

\subsection*{$^{207}$Bi}
Germain observed the decay of $^{207}$ Bi in 1951 as described in ``Auger effect in astatine'' \cite{1950Ger01}. A bismuth target was bombarded with a 30~MeV $^4$He beam from the Berkeley 184-inch cyclotron forming $^{211}$At in the ($\alpha$,2n) reaction. $^{207}$Bi was then populated by $\alpha$ decay. Auger electrons were measured with photographic plates following chemical separation. ``After 10 days, the activities of At$^{210}$ and At$^{211}$ would be only 10$^{-10}$ of the original activities. Thus, these Auger electrons must have come from a daughter activity. The possible daughters are Pb$^{207}$, Po$^{210}$, and Bi$^{207}$. It is known that Pb$^{207}$ is stable and Po$^{210}$ is an alpha-emitter. Therefore, these Auger electrons in all probability have come from the K-capture decay of Bi$^{207}$. Furthermore, the only alpha-particles found on the plate were those of Po$^{210}$ (these coming from the decay of At$^{210}$). Therefore, one can conclude that Bi$^{207}$ decays by K-capture and its half-life is very long, probably several years.'' The currently accepted half-life is 32.9(17)~y. The previous assignment of a 6.4~d half-life to $^{207}$Bi \cite{1941Faj01,1941Faj02} was evidently incorrect.

\subsection*{$^{208}$Bi}
In the 1953 article ``Energy levels in lead and bismuth and nuclear shell structure'' Harvey reported the observation of $^{208}$Bi \cite{1953Har01}. A 15.5 MeV deuteron beam from the MIT cyclotron bombarded a bismuth target and $^{208}$Bi was populated in (d,t) reactions. Triton spectra were measured with a triple coincidence proportional counter. ``The three levels observed in Bi$^{208}$ arise from the p$_{1/2}$, f$_{5/2}$, and p$_{3/2}$ neutrons interacting with the odd h$_{9/2}$ proton.'' Earlier, Krishnan had assigned a 6.35(20)~d half-life incorrectly to $^{208}$Bi \cite{1940Kri01}.

\subsection*{$^{209}$Bi}
In 1924 Aston reported the observation of $^{209}$Bi in ``The mass-spectra of cadmium, tellurium, and bismuth'' \cite{1924Ast03}. Metallic bismuth was ground in the anode mixture of the Cavendish mass spectrometer. ``This hope was realized with an anode containing metallic bismuth, and a single line appeared in the expected position $-$ 209. This line is very faint, and owing to the great mass lies in an unfavorable part of the plate, but there seems no reason to doubt that bismuth is a simple element of mass number 209, as recent determinations of its atomic weight suggest.''

\subsection*{$^{210}$Bi}
Rutherford reported on a new activity later assigned to $^{210}$Bi in ``Slow transformation products of radium'' in 1905 \cite{1905Rut02}. Alpha- and beta-ray decay spectra of slow transformation products separated physically and chemically from fast decaying radium were measured. ``I have recently examined more carefully the product radium D, and have found strong evidence that it is not a single product, but contains two distinct substances. The
parent product, radium D, does not give out rays at all, but changes into a substance which gives out only $\beta$ rays, and is half transformed in about six days. Unless observations are made on the product radium D shortly after its separation, this rapid change is likely to escape detection. The work on this subject is still in progress, but the evidence at present obtained indicates that the active deposit from the emanation, after passing through the three rapid stages, represented by radium A, B, and C, is transformed into a `rayless' product D, which changes extremely slowly. D continuously produces from itself another substance - which may for the time be termed D$_1$ - which is transformed in the course of a few weeks and emits only $\beta$ rays.'' The reported half-life for radium D$_1$ ($^{210}$Bi) of about six days is consistent with the presently adopted value of 5.012(5)~d. As stated in the quote Rutherford had earlier missed this activity, assigning radium D as a pure $\beta$ emitter with a half-life of about 40~years and radium E as an $\alpha$ emitter with a half-life of about 1~year \cite{1904Rut02,1905Rut01}. In a paper submitted four months later Rutherford renamed radium D$_1$ to radium E and the old radium E to radium F \cite{1905Rut03}. In October 1904, Hofmann et al.\ \cite{1904Hof01} had identified a $\beta$-emitter chemically separated from radiolead (radium D, $^{210}$Pb) which Rutherford acknowledges \cite{1905Rut03}, however, the extracted half-life of about six weeks was incorrect.

\subsection*{$^{211,212}$Bi}
Rutherford reported two activities later identified as $^{211}$Bi and $^{212}$Bi in the 1905 paper ``Bakerian lecture.-The succession of changes in radioactive bodies'' \cite{1905Rut01}. The decay curves of ``excited activities'' following the decay of actinium emanation ($^{219}$Rn) and thorium emanation ($^{220}$Rn) were measured.  ``We may thus conclude that the active deposit from actinium undergoes two distinct successive transformations: (1) A rayless change, in which half the matter is transformed in 41 minutes; (2) A change giving rise to $\alpha$ rays, in which half the matter is transformed in 1.5 minutes.'' For the thorium activities Rutherford states: ``The evidence, as a whole, thus supports the conclusion that the active deposit from thorium undergoes two successive transformations as follows: (1) A `rayless' change for which $\lambda_1$ = 1.75$\times$10$^{-5}$, i.e., in which half the matter is transformed in 11 hours; (2) A second change giving rise to $\alpha$, $\beta$ and $\gamma$ rays, for which $\lambda_2$ = 2.08$\times$10$^{-4}$, i.e., in which half the matter is transformed in 55 minutes.'' The measured half-lives for the second decay of the actinium emanation (1.5~min, named actinium B, $^{211}$Bi) and the second decay of the thorium emanation (55~min, named thorium B) agree with the presently adopted values of 2.14(2)~min for $^{211}$Bi and 60.55(6)~min for $^{212}$Bi, respectively. Rutherford and Soddy had reported the observation of a ``Thorium-excited activity II'' from the decay of thorium emanation earlier \cite{1903Rut01}. ThB and AcB were later reclassified as ThC and AcC, respectively.

\subsection*{$^{213}$Bi}
Hagemann et al.\ discovered $^{213}$Bi in 1947 in ``The (4n+1) radioactive series: the decay products of U$^{233}$'' \cite{1947Hag01}. The half-lives and $\alpha$- and $\beta$-decay energies of the nuclides in the decay chain of $^{233}$U were measured. ``These decay products, which constitute a substantial fraction of the entire missing, 4n+1, radioactive series are listed together with their radioactive properties, in [the table].'' The measured half-life of 47 min agrees with the presently accepted value of 45.59(6)~min. Hagemann et al.\ acknowledge the simultaneous observation of $^{213}$Bi by English et al.\ which was submitted only a day later and published in the same issue of Physical Review on the next page \cite{1947Eng01}.

\subsection*{$^{214}$Bi}
In the 1904 paper ``Heating effect of the radium emanation'', Rutherford and Barnes described the discovery of a new activity later identified as $^{214}$Bi \cite{1904Rut01}. The decay curves of ``excited activities'' following the decay of radium and radium emanation ($^{222}$Rn) were measured. ``An analysis of the decay curves of excited activity, produced for different intervals of exposure in the presence of the emanation, shows that there are three well-marked changes occurring in emanation X of radium. In the first change, half the matter is transformed in 3 minutes; in the second, half in 34 minutes; and in the third, half in 28 minutes. A full account of the analysis of these changes and their peculiarities will be given by one of us in a later paper. The first change is accompanied only by $\alpha$ rays, the second change is not accompanied by $\alpha$ rays at all, and the third change by $\alpha$, $\beta$, and $\gamma$ rays.'' The 28~min half-life of the third decay - later named radium C - is close to the currently adopted half-life of 19.9(4)~min. Rutherford and Soddy had reported the observation of a ``Radium-excited activity III'' from the decay of radium emanation earlier \cite{1903Rut01}.

\subsection*{$^{215}$Bi}
In 1953 $^{215}$Bi was first reported by Hyde and Ghiorso in ``The alpha-branching of AcK and the presence of astatine in nature'' \cite{1953Hyd01}. A 20-mC $^{227}$Ac source was used to study the nuclide of the 4n+3 decay series by chemical and physical separation and measuring the radioactivity with an alpha-ray differential pulse analyzer. ``The observed branching rate is ca 4$\times$10$^{-5}$, and the At$^{219}$ daughter decays predominantly by the emission of 6.27 Mev alpha-particles with a half-life of 0.9 minute to the new isotopes Bi$^{215}$, which in turn emits $\beta^-$ particles with a half-life of 8 minutes.'' This half-life agrees with the currently adopted value of 7.6(2)~min.

\subsection*{$^{216}$Bi}
In the 1989 paper ``New neutron-rich isotopes of astatine and bismuth,'' Burke et al.\ described the observation of $^{216}$Bi \cite{1989Bur01}. A thorium/tantalum metal-foil target was bombarded with 600~MeV protons from the CERN synchro-cyclotron. Astatine isotopes were produced in spallation reactions and separated with the ISOLDE-II on-line separator. $^{216}$Bi was observed from the $\alpha$-decay of $^{220}$At. ``The 8\% alpha branch of $^{220}$At feeds the daughter nucleus $^{216}$Bi, the half-life of which has been determined to be 6.6(21) min.'' This half-life corresponds to an isomeric state. The ground state was measured for the first time eleven years later \cite{2000Kur01}.

\subsection*{$^{217,218}$Bi}
Pf\"utzner et al.\ observed $^{217}$Bi and $^{218}$Bi as described in the 1998 publication ``New isotopes and isomers produced by the fragmentation of $^{238}$U at 1000 MeV/nucleon'' \cite{1998Pfu01}. A 1~GeV/nucleon $^{238}$U beam from the SIS facility at GSI bombarded a beryllium target and projectile fragments were identified with the fragment separator FRS in the standard achromatic mode. ``The nuclei $^{209}$Hg, $^{210}$Hg, $^{211}$Tl, $^{212}$Tl, $^{218}$Bi, $^{219}$Po and $^{220}$Po have been identified for the first time.'' The observation of $^{217}$Bi was not considered a discovery quoting a paper by Van Duppen et al.:``A complementary technique is the combination of high-energy proton-induced spallation of thick heavy targets with on-line isotope separation (ISOL), with which the isotopes $^{215}$Pb and $^{217}$Bi were recently observed \cite{1998Van01}.'' However, Van Duppen et al.\ did not actually observe $^{217}$Bi stating: ```In conclusion, we have presented a new method that allows detailed decay-spectroscopy studies of the neutron-rich `east of $^{208}$Pb' using the pulsed release from the ISOLDE targets... It has been successfully applied in a recent experiment where two new isotopes ($^{215}$Pb and $^{217}$Bi) were identified ...'' and referring to ``K. Rykaczewski {\it et al.}, to be published.'' This clearly does not constitute the discovery of $^{217}$Bi, thus the credit is given to Pf\"utzner et al.\ because $^{217}$Bi is cleanly identified in the mass-to-charge spectra.


\subsection*{$^{219-223}$Bi}
In 2009 Alvarez-Pol et al.\ reported the observation of $^{219}$Bi, $^{220}$Bi, $^{221}$Bi, $^{222}$Bi, and $^{223}$Bi in the paper ``Production cross-sections of neutron-rich Pb and Bi isotopes in the fragmentation of $^{238}$U'' \cite{2009Alv01}. The GSI SchwerIonen Synchrotron SIS was used to bombard a beryllium target with a 1 A GeV $^{238}$U beam. Reaction products were identified by mass and atomic number based on magnetic rigidity, time of flight, and energy loss. ``The high-resolving power FRagment Separator (FRS) at GSI allowed us to identify and determine the production cross-sections of 43 nuclei, nine of them for the first time ($^{216,217,218,219}$Pb and $^{219,220,221,222,223}$Bi).''

\subsection*{$^{224}$Bi}
$^{224}$Bi was discovered by Alvarez-Pol and the results were published in the 2010 paper ``Production of new neutron-rich isotopes of heavy elements in fragmentation reactions of $^{238}$U projectiles at 1A GeV'' \cite{2010Alv01}. A beryllium  target was bombarded with a 1~A GeV $^{238}$U beam from the GSI SIS synchrotron. The isotopes were separated and identified with the high-resolving-power magnetic spectrometer FRS. ``To search for new heavy neutron-rich nuclei, we tuned the FRS magnets for centering the nuclei $^{227}$At, $^{229}$At, $^{216}$Pb, $^{219}$Pb, and $^{210}$Au along its central trajectory. Combining the signals recorded in these settings of the FRS and using the analysis technique previously explained, we were able to identify 40 new neutron-rich nuclei with atomic numbers between Z=78 and Z=87; $^{205}$Pt, $^{207-210}$Au, $^{211-216}$Hg, $^{214-217}$Tl, $^{215-220}$Pb, $^{219-224}$Bi, $^{223-227}$Po, $^{225-229}$At, $^{230,231}$Rn, and $^{233}$Fr.'' It is interesting to note that although the paper claims the discovery of additional bismuth isotopes, $^{219-223}$Bi, the observation of these isotopes had been reported earlier by the same authors \cite{2009Alv01}.


\section{$^{186-227}$Po}\vspace{0.0cm}
The discovery of the element polonium coincided with the discovery of the radioactivity due to $^{210}$Po. In 1898 M. Curie suggested the presence of a new element by observing strong activities in pitchblende and chalcolite \cite{1898Cur01}. Later in the same year, P. Curie and M. Curie presented evidence that they had observed a new element and suggested the name polonium \cite{1898Cur03}. Marckwald reported the discovery of a new substance he called radio tellurium in 1902 \cite{1902Mar02,1902Mar03}. In 1904/1905 Rutherford showed that polonium and radio-tellurium were they same substance as what he identified as radium F \cite{1904Rut02,1905Rut02,1905Rut03}.

\begin{figure}
	\centering
	\includegraphics[scale=.7]{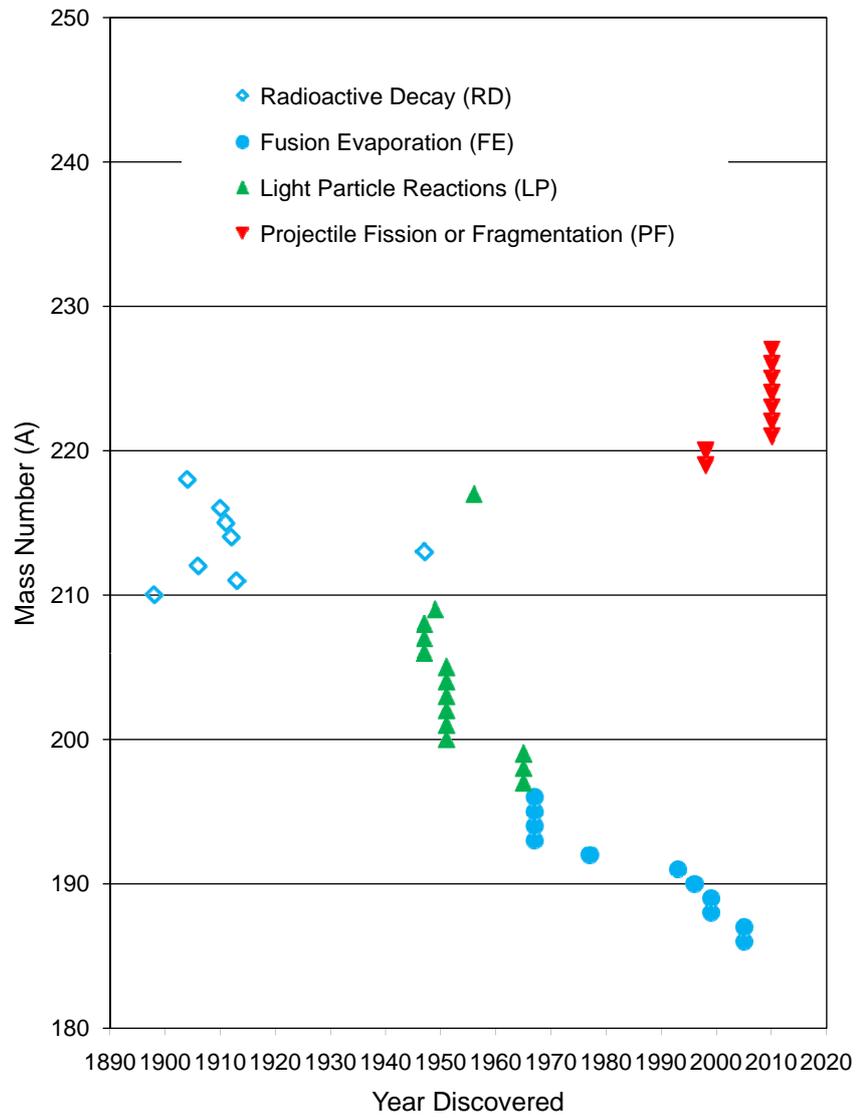}
	\caption{Polonium isotopes as a function of time when they were discovered. The different production methods are indicated.}
\label{f:year-polonium}
\end{figure}

Forty-two polonium isotopes from A = 186--227 have been discovered so far. According to the HFB-14 model \cite{2007Gor01} about 45 additional polonium isotopes could exist. Figure \ref{f:year-polonium} summarizes the year of first discovery for all polonium isotopes identified by the method of discovery: radioactive decay (RD), fusion evaporation reactions (FE), light-particle reactions (LP), and projectile fission or fragmentation (PF). In the following, the discovery of each polonium isotope is discussed in detail and a summary is presented in Table 1.

\subsection*{$^{186,187}$Po}
Andreyev et al.\ announced the discovery of $^{186}$Po and $^{187}$Po in the 2005 article ``Cross section systematics for the lightest Bi and Po nuclei produced in complete fusion reactions with heavy ions'' \cite{2005And01}. $^{46}$Ti beams of 202$-$242~MeV from the GSI UNILAC bombarded a SmF$_3$ target enriched in $^{144}$Sm forming $^{186}$Po and $^{187}$Po in (4n) and (3n) fusion-evaporation reactions, respectively. Reaction products were separated with the velocity filter SHIP and implanted into a position-sensitive silicon detector which also measured subsequent $\alpha$ and proton emission. $^{186}$Po and $^{187}$Po were identified with the method of genetically correlated events. ``For the sake of completeness, in [the table] we also provide preliminary cross section values for the new isotopes $^{186,187}$Po, produced recently at SHIP in, respectively, the 4n and 3n evaporation channels of the complete fusion reaction $^{46}$Ti$+^{144}$Sm$\rightarrow^{190}$Po$^*$.'' Andreyev et al.\ refer to a paper ``to be published'' for further details, however, no details for $^{187}$Po were published in a later paper \cite{2006And01}.

\subsection*{$^{188,189}$Po}
In 1999, Andreyev et al.\ reported the first observation of $^{188}$Po and $^{189}$Po in the paper ``Alpha decay of the new isotopes $^{188,189}$Po'' \cite{1999And01}. A $^{142}$Nd target was bombarded with 239$-$307~MeV $^{52}$Cr beams from the GSI UNILAC heavy ion accelerator producing $^{188}$Po and $^{189}$Po in (6n) and (5n) fusion-evaporation reactions, respectively. Recoils were separated with the velocity filter SHIP and implanted in a 16-strip position-sensitive silicon detector which also recorded subsequent $\alpha$ decays. ``On the basis of these data we assign the 7915(25)~keV line to the $\alpha$ decay of $^{188}$Po... The new isotope $^{189}$Po was identified by observing nineteen 7264(15)~keV-280(1)~keV $\alpha$-$\gamma$ events in [the figure] and 78(9) 7264(25)~keV-e$^-$ $\alpha$-e$^-$ coincidences in [the figure]. The half-life values of these two groups of events, deduced from the time interval between an implant and the $\alpha$-$\gamma$ or $\alpha$-e$^-$ coincidence pair are very similar and the combined data give T$_{1/2}$ = 5(1)~ms.'' The measured half-life of 400$^{+200}_{-150}$~ms for $^{188}$Po corresponds to the presently adopted value. For $^{189}$Po, the measured 5(1)~ms half-life agrees with the current value of 3.5(5)~ms.

\subsection*{$^{190}$Po}
In the 1996 paper ``Determination of the $^{190}$Po $\alpha$ reduced width'' Batchelder et al.\ announced the discovery of $^{190}$Po \cite{1996Bat01}. An enriched $^{144}$Sm target was bombarded with a 215~MeV $^{48}$Ti beam from the Berkeley 88-inch cyclotron forming $^{190}$Po in (2n) fusion-evaporation reactions. Recoils were stopped in aluminum catcher foils and rotated in front of an array of Si detectors. ``The isotope $^{190}$Po was produced in the $^{144}$Sm($^{48}$Ti,2n) reaction and its $\alpha$-decay energy and half-life were measured to be 7.49(4)~MeV and 2.0$^{+0.5}_{-1.0}$~ms, respectively.'' The quoted half-life agrees with the currently accepted value of 2.46(5)~ms.

\subsection*{$^{191}$Po}
Quint et al.\ first observed $^{191}$Po in 1993 and reported their results in ``Investigation of the fusion of heavy nearly symmetric systems'' \cite{1993Qui01}. An enriched $^{100}$Mo target was bombarded with a 3$-$5~MeV per nucleon $^{94}$Mo beam from the GSI UNILAC forming $^{191}$Po in (3n) fusion-evaporation reaction. Recoil products were separated with the velocity filter SHIP and implanted into a surface-barrier detector which also measured subsequent $\alpha$ decay. ``To demonstrate the sensitivity of this method [the figure] shows an $\alpha$ spectrum which includes a peak which we assign to $^{191}$Po, an isotope which has been seen here for the first time, to our knowledge, and which has been produced in the reaction $^{100}$Mo($^{94}$Mo,3n).'' The measured half-life of 15.5($^{+6}_{-2.5}$)~ms agrees with the currently adopted value 22(1)~ms.

\subsection*{$^{192}$Po}
In the 1977 article ``Very neutron-deficient polonium isotopes produced through $^{20}$Ne induced reactions,'' della Negra et al.\ reported the observation of $^{192}$Po \cite{1977Del01}. An enriched $^{182}$W target was bombarded with a 10~MeV/amu $^{20}$Ne beam from the Orsay ALICE accelerator. Reaction products were transported with a helium jet to a silicon detector which measured $\alpha$ particles. Mass assignments were made by measuring excitation functions. ``The alpha activity observed at an energy of 7.12~MeV is attributed to mass 192. Because of the low intensity the half life has not been measured.'' The currently adopted half-life is 33.2(14)~ms. An earlier report of a 6.58(4)~MeV $\alpha$ activity with a 0.5(1)~s half-life \cite{1958Tov01} was evidently incorrect.

\subsection*{$^{193-196}$Po}
The 1967 paper ``$^{193-200}$Po isotopes produced through heavy ion bombardments''  by Siivola discussed the observation of $^{193}$Po, $^{194}$Po, $^{195}$Po, and $^{196}$Po \cite{1967Sii01}. Enriched $^{185}$Re targets were bombarded with 150$-$185~MeV $^{19}$F beams from the Berkeley linear accelerator forming $^{193-196}$Po in (11n-8n) fusion-evaporation reactions. Recoils were deposited on an aluminum plate with a helium jet and subsequent $\alpha$ spectra were measured with a solid state counter. ``$^{196}$Po: A 6.526$\pm$0.008~MeV 5.5$\pm$0.5~s activity observed in $^{19}$F+$^{185}$Re bombardments is assigned to this isotope... $^{195}$Po: In $^{19}$F+$^{185}$Re bombardments a pair of alpha groups was present with energies of 6.624$\pm$0.008~MeV and 6.710$\pm$0.010~MeV and half-lives of 4.5$\pm$0.5~s and 2.0$\pm$0.2~s... $^{194}$Po: The last of the activities whose half-life was measured is a 6.847$\pm$0.010~MeV, 0.6$\pm$0.2~s activity... $^{193}$Po: The highest alpha energy observed is 6.98$\pm$0.02~MeV.'' The measured values of 0.6(2)~s for $^{194}$Po, 4.5(5)~s for $^{195}$Po, and 5.5(5)~s for $^{196}$Po, agree with the currently accepted values of 0.392(4)~s, 4.64(9)~s, and 5.8(2)~s, respectively. Earlier measurements assigning significantly longer half-lives to $^{193-195}$Po \cite{1958Tov01}, and $^{196}$Po \cite{1959Att01,1964Bru01} were evidently incorrect.

\subsection*{$^{197-199}$Po}
Brun et al.\ reported the observation of $^{197}$Po, $^{198}$Po, and $^{199}$Po in the 1965 paper ``Caract\'eristiques des d\'esint\'egrations alpha des isotopes l\'egers du polonium'' \cite{1965Bru01}. Bismuth targets were irradiated with 80$-$155~MeV protons from the Orsay synchrocyclotron forming $^{197}$Po, $^{198}$Po, and $^{199}$Po in (p,xn) reactions. Excitation functions and $\alpha$ spectra were measured following chemical separation. ``Pour l'isotope 199 nous avons d\^u admettre deux isom\`eres (T = 4.1~m pour $\alpha$ de 6.04~MeV et T = 5.5~m pour $\alpha$ de 5.93~MeV). La raie alpha de 6.16~MeV attribu\'ee autrefois \`a $^{196}$Po est due \`a une d\'esint\'egration de $^{198}$Po (T = 1.7~m) et enfin pour $^{197}$Po deux isom\`eres ont \'et\'e trouv\'es, l'un de T = 58~s (alpha de 6.27~MeV environ) et l'autre de T = 29 pour lequel la raie alpha de 6.37~MeV est au moins trois fois plus intense.'' [For the isotope 199 we observed two isomers (T$_{1/2}$ = 4.1~min for a 6.4~MeV $\alpha$ and T$_{1/2}$ = 5.5~min for a 5.93~MeV $\alpha$). The 6.16~MeV $\alpha$ line formerly attributed to $^{196}$Po is due to the decay of $^{198}$Po (T$_{1/2}$ = 1.7~m) and finally two isomeric states were found in $^{197}$Po, one with T$_{1/2}$ = 58~s (6.27~MeV $\alpha$) and the other with T$_{1/2}$ = 29~s with a 6.37~MeV $\alpha$ which is at least three times more intense.] These measured half-lives of 58~s for $^{197}$Po, 1.7~m for $^{198}$Po, and 5.2~m for $^{199}$Po, agree with the currently accepted values of 53.6(10)~s, 1.77(3)~m, and 5.48(16)~min, respectively. The group had reported half-lives of these polonium isotopes a year earlier \cite{1964Bru01}, however, the mass assignments were incorrect. Also, tentative assignments of 4~min to $^{197}$Po \cite{1954Ros02,1959Att01,1959Att02}, 6~min \cite{1954Ros02} and $\sim$7~min \cite{1959Att02} to $^{198}$Po and 11~min \cite{1954Ros02}, 12~min \cite{1959Att01,1959Att02} and 13(3)~min \cite{1961For02} to $^{199}$Po were evidently incorrect.

\subsection*{$^{200,201}$Po}
In 1951, Karraker et al.\ announced the discovery of $^{200}$Po and $^{201}$Po in the paper ``Alpha-decay energies of polonium isotopes'' \cite{1951Kar02}. Bismuth oxide was bombarded with 150 MeV protons from the Berkeley 60-inch and 184-inch cyclotrons. Alpha-decay spectra were measured with an ionization chamber following chemical separation. ``The isotopes Po$^{200}$ and Po$^{201}$ have been shown to have half-lives of 11 and 18~minutes, respectively, and to emit alpha-particles of 5.84 and 5.70~Mev.'' These values are close to the currently accepted half-lives of 11.51(8)~min and 15.6(1)~min for $^{200}$Po and $^{201}$Po, respectively.

\subsection*{$^{202-205}$Po}
$^{202}$Po, $^{203}$Po, $^{204}$Po, and $^{205}$Po, were discovered in 1951 and reported in  ``Polonium isotopes produced with high energy particles'' by Karraker and Templeton \cite{1951Kar01}. Natural lead and bismuth targets were bombarded with helium beams and protons and deuterons, respectively, from the 184-in Berkeley cyclotron. Decay curves were measured following chemical separation. ``52-min Po$^{202}$ and 95-min Bi$^{202}$: In addition to the other bismuth activities mentioned above, we observed a 95-minute activity among the bismuth daughters of polonium produced in bombardments at fairly high energy... 47-min Po$^{203}$: ...The data indicated another polonium isotopes of a shorter half-life than Po$^{204}$. It must be at mass 203, since 52-hr Pb$^{203}$ appeared in the decay curves of the first bismuth fractions, but not in the later ones... 3.8-hr Po$^{204}$: ...These experiments have fixed the 3.8-hour polonium at mass 204... 1.5-hr Po$^{205}$:... These data confirm the assignment of the 1.5-hour polonium to mass 205.'' The half-lives of 52~min ($^{202}$Po), 47(5)~min ($^{203}$Po), 3.8~h($^{204}$Po), and 1.5~h ($^{205}$Po) agree with the currently accepted values of 44.6(4)~min, 36.7(5)~min, 3.53(2)~h, and 1.74(8)~h, respectively.

\subsection*{$^{206-208}$Po}
In the 1947 paper ``Artificial radioactive isotopes of polonium'', Templeton et al.\ reported the observation of $^{206}$Po, $^{207}$Po, and $^{208}$Po \cite{1947Tem01}. Enriched $^{204}$Pb, $^{206}$Pb, and $^{207}$Pb were bombarded with a 40~MeV $^4$He beam from the Berkeley 60-inch cyclotron populating $^{206}$Po, $^{207}$Po, and $^{208}$Po in ($\alpha$,2n), ($\alpha$,3n), and ($\alpha$,3n), respectively. $^{208}$Po was also produced by bombarding bismuth targets with protons and deuterons. Products were chemically separated and electrons and electromagnetic radiation were measured with Geiger tubes. ``The 9-day polonium is 206 or lighter, since it is made in good yield from Pb$^{204}$. But its daughter the 6.4-d.\ Bi, is 206 or 207. Therefore both must be 206, which is confirmed by other arguments stated above for the bismuth isotope... 5.7-Hour Po$^{207}$ ...Yield data presented below for different isotopic mixtures show that the 5.7-hour isotope is made chiefly from Pb$^{206}$ by 40-Mev helium ions. It is not made in appreciable yield from Bi$^{209}$ by 20-Mev deuterons. If the ($\alpha$,3n) and (d,3n) reactions are prolific while the ($\alpha$,4n) and (d,4n) are not at these energies, the isotopic assignment must be Po$^{207}$... 3-Year Po$^{208}$: ...Direct decay measurements for ten months, corrected for Po$^{210}$ from pulse-analysis data, showed decay of the new polonium, isotope with about a 3-year half-life. A large sample of this isotope after five months had so little Geiger activity associated with it, if sufficient mica was interposed to stop the alpha-particles, that the half-life of the lead daughter, if it emits x-rays, must be longer than 100 years. From arguments presented below, based on yields from the different isotopic mixtures, the isotope is assigned to Po$^{208}$.'' The half-lives of 9~d ($^{206}$Po), 5.7~h ($^{207}$Po), and 3-y ($^{208}$Po) agree with the currently accepted values of 8.8(1)~d, 5.80(2)~h, and 2.898(2)~y, respectively.

\subsection*{$^{209}$Po}
Kelly and Segre first observed $^{209}$Po and reported their results in the 1949 paper ``Some excitation functions of bismuth'' \cite{1949Kel01}. Bismuth targets were bombarded with 19~MeV deuterons from the Berkeley 60-inch cyclotron. Resulting activities were measured with a parallel plate ionization chamber. ``The Po$^{208}$ activity seemed to be produced by deuterons of an energy too low to make a (d,3n) reaction. A further close examination of the alpha-activity in the energy region between 10 and 15 Mev showed that we had also another Po isotope present, emitting alphas of 4.95~Mev. This substance is Po$^{209}$ formed by the (d,2n) reaction. If we assume that it decays only by $\alpha$-emission and that the maximum cross section for its formation is about 10$^{-24}$~cm$^2$, a half-life of about 200~years results.'' The currently accepted half-life is 102(5)~y.

\subsection*{$^{210}$Po}
In 1898 P. Curie and M. Curie reported the observation of a new radioactive element in ``Sur une substance nouvelle radio-active, contenue dans la pechblende'' \cite{1898Cur03}. They separated a radioactive substance from pitchblende which was 400 times more radioactive than uranium. ``Nous croyons donc que la substance que nous avons retir\'ee de la pechblende contient un m\'etal non encore signal\'e, voisin du bismuth par ses propri\`et\'es analytiques. Si l'existence de ce nouveau m\'etal se confirme, nous proposons de l'appeler polonium, du nom du pays d'origine de l'un de nous.'' [We therefore believe that the substance we have extracted from pitchblende contains a metal that has not been observed before, with analytical properties similar to bismuth. If the existence of this new metal is confirmed, we suggest to call it polonium, named after the country of origin of one of us.] Earlier M. Curie had suggested the presence of a new element by observing strong activities in pitchblende and chalcolite \cite{1898Cur01}. In 1904/1905 Rutherford showed that the polonium activity, radio-tellurium (discovered in 1902 by Marckwald \cite{1902Mar02,1902Mar03}) and radium F (originally named radium E) were the same ($^{210}$Po) \cite{1904Rut02,1905Rut02,1905Rut03}.

\subsection*{$^{211}$Po}
In 1913 Marsden and Wilson discovered a new activity from actinium C as described in ``Branch product in actinium C'' \cite{1913Mar01}. An active source of actinium was covered with a sheet of mica and placed in a chamber. The number of scintillations on a zinc sulphite screen as a function of pressure in the chamber were counted. ``The results showed that in addition to the $\alpha$ particles of actinium C with a range of 5.4~cm., a small number, about 1 in 600, can penetrate as far as about 6.45~cm. Special experiments showed that the long-range $\alpha$ particles could not be due to radium or thorium impurity, and they must therefore be attributed to the expected new branch product.'' This activity later named actinium C' corresponds to $^{211}$Po.

\subsection*{$^{212}$Po}
In ``\"Uber einige Eigenschaften der $\alpha$-Strahlen des Radiothoriums. I.'', a new actitivty from thorium emanation was reported by Hahn in 1906 \cite{1906Hah02}. Radium B sources were separated from thorium emanation and the ionization was measured as a function of distance. ``Die komplexe Kurve zeigt also an, da\ss\ der aktive Beschlag an dem Draht zwei verschiedene Arten von $\alpha$-Partikeln mit verschiedenem Durchdringungsverm\"ogen f\"ur Luft aussendet, also aus zwei verschiedenen $\alpha$-Strahlenprodukten besteht... Der von Rutherford gew\"ahlten Nomenklatur folgend, wird man das neue $\alpha$-Produkt des Thoriums `Thorium C' zu nennen haben,'' [The complex curve demonstrates that the active substance at the wire emits two kinds of $\alpha$-particles with different ranges in air, thus consisting of two different $\alpha$-sources... According to Rutherford's convention, the new $\alpha$-product of thorium should be named ``thorium C''.] Later it was renamed to thorium C' corresponding to $^{212}$Po.

\subsection*{$^{213}$Po}
Hagemann et al.\ discovered $^{213}$Po in 1947 in ``The (4n+1) radioactive series: the decay products of U$^{233}$'' \cite{1947Hag01}. The half-lives and $\alpha$- and $\beta$-decay energies of the nuclides in the decay chain of $^{233}$U were measured. ``These decay products, which constitute a substantial fraction of the entire missing, 4n+1, radioactive series are listed together with their radioactive properties, in [the table].'' Only the decay energy was measured and the half-life is listed as ``very short''. Hagemann et al.\ acknowledge the simultaneous observation of $^{213}$Po by English et al.\ which was submitted only a day later and published in the same issue of Physical Review on the next page \cite{1947Eng01}.

\subsection*{$^{214}$Po}
Fajans reported a new $\alpha$ activity in the decay of radium C in the 1912 article ``\"Uber die Verzweigung der Radiumzerfallsreihe'' \cite{1912Faj01}. Recoil products from radium C and radium B were studied and $\alpha$- and $\beta$-decay rates and relative intensities were measured. ``Um jetzt ein vollst\"andiges Bild der im RaC vor sich gehenden Umwandlungen zu gewinnen, mu\ss\ ber\"ucksichtigt werden, da\ss\ es auf Grund der Untersuchung von H.\ Geiger und J.\ M.\ Nuttall nicht m\"oglich ist, da\ss\ die $\alpha$-Strahlen des RaC von der Reichweite 7,1~cm einem Produkt wie RaC mit der Halbwertszeit von 19~1/2 Minuten zukommen, und da\ss\ es also im RaC au\ss er RaC$_1$ und RaC$_2$ noch ein drittes diese $\alpha$-Strahlen lieferndes Produkt mit einer Periode von etwa 10$^{-6}$ Sekunden geben mu\ss.'' [In order to gain a complete picture of the RaC tranformations, it is necessary to consider that it is not possible - due to the studies by H.\ Geiger and J.\ M.\ Nuttall - that the $\alpha$-rays with a range of 7.1~cm are coming from a product like RaC with a half-life 19~1/2 minutes. Thus, in addition to RaC$_1$ and RaC$_2$, a third product which produces these $\alpha$-rays with a half-life of about 10$^{-6}$~s must exist in RaC.] Fajans named this product RaC' which corresponds to $^{214}$Po.

\subsection*{$^{215}$Po}
A new $\alpha$ activity in the actinium decay chain was identified in 1911 by Geiger as reported in ``The transformation of the actinium emanation'' \cite{1911Gei01}. Scintillations were counted using a zinc sulfide screen fixed to a microscope. The range of the $\alpha$ particles was measured by changing the distance in air. Also, a box was constructed that allowed to calculate the half-life from the number of scintillations as a function of an applied electric field. ``It has been shown that the actinium emanation is complex, consisting of two products each of which emits $\alpha$ rays. The first one - the emanation - with a period of 4 seconds emits $\alpha$ rays of a range of 5.7~cm. The second product emits $\alpha$ rays of range 6.5~cm., and has a period of the order of 1/500 of a second.'' This $\alpha$-emitter was later named AcA \cite{1911Rut01} corresponding to $^{215}$Po. This half-life agrees with the presently accepted value of 1.781(4)~ms. Earlier Geiger and Marsden had established an upper half-life limit of 0.1~s for this activity \cite{1910Gei02}.

\subsection*{$^{216}$Po}
In 1910 Geiger and Marsden reported the observation of short lived $\alpha$-activity in the decay of thorium emanation: ``\"Uber die Zahl der von der Aktinium- und Thoriumemanation ausgesandten $\alpha$-Teilchen'' \cite{1910Gei02}. A thorium emanation source was placed between two zinc sulphide screens which were viewed by two microscopes. The number of simultaneous or emissions in short successions were counted. ``Derartige Szintillationen in rascher Aufeinanderfolge zeigten sich auch dann, wenn die pro Minute beobachtete Zahl von Szintillationen \"au\ss erts gering war, so da\ss\ die Wahrscheinlichkeit eines zuf\"alligen zeitlichen und \"ortlichen Zusammentreffens zweier Szintillationen verschwindend klein ist. Es erscheint damit die Existenz von mindestens einem $\alpha$-Strahlenprodukt in der Thoriumemanationsgruppe von mittlerer Lebensdauer von etwa 1/5 Sekunde erwiesen.'' [This rapid succession of scintillations were also present, when the observed number of scintillations per minute were extremely small, so that the probability for random temporal and spacial coincidences of two scintillations were vanishingly small. Therefore it seems that the existence of at least one $\alpha$-decay product in the thorium emanation group with a half-life of about 1/5 seconds has been demonstrated.] This activity was later named ThA \cite{1911Rut01} and corresponds to $^{216}$Po. The half-life agrees with the currently adopted value of 0.145(2)~s.

\subsection*{$^{217}$Po}
Momyer and Hyde reported the observation of $^{217}$Po in the 1956 paper ``Properties of Em$^{221}$'' \cite{1956Mom01}. Thorium targets were bombarded with 110~MeV protons from the 184-inch Berkeley cyclotron. Alpha-decay spectra were measured following chemical separation. ``In 20 percent of its disintegrations Em$^{221}$ emits an alpha particle of 6.0-Mev energy unresolved from the 6.0-Mev alpha particle of Fr$^{221}$, and gives rise to Po$^{217}$, a previously unreported isotope of polonium. Po$^{217}$ has a half-life of less than 10 seconds and emits alpha particles of 6.54$\pm$0.02~Mev.'' The currently accepted half-life is 1.46(5)~s.

\subsection*{$^{218}$Po}
In the 1904 paper ``Heating effect of the radium emanation'', Rutherford and Barnes described the discovery of a new activity later identified as $^{218}$Po \cite{1904Rut01}. The decay curves of ``excited activities'' following the decay of radium and radium emanation ($^{222}$Rn) were measured. ``An analysis of the decay curves of excited activity, produced for different intervals of exposure in the presence of the emanation, shows that there are three well-marked changes occurring in emanation X of radium. In the first change, half the matter is transformed in 3 minutes; in the second, half in 34 minutes; and in the third, half in 28 minutes. A full account of the analysis of these changes and their peculiarities will be given by one of us in a later paper. The first change is accompanied only by $\alpha$ rays, the second change is not accompanied by $\alpha$ rays at all, and the third change by $\alpha$, $\beta$, and $\gamma$ rays.'' The 3~min half-life of the first decay - later named radium A - agrees with the currently adopted half-life of 3.098(12)~min. Rutherford and Soddy had reported the observation of a ``Radium-Excited Activity I'' from the decay of radium emanation earlier \cite{1903Rut01}.

\subsection*{$^{219,220}$Po}
Pf\"utzner et al.\ reported the discovery of $^{219}$Po and $^{220}$Po in the 1998 publication ``New isotopes and isomers produced by the fragmentation of $^{238}$U at 1000 MeV/nucleon'' \cite{1998Pfu01}. A 1~GeV/nucleon $^{238}$U beam from the SIS facility at GSI bombarded a beryllium target and projectile fragments were identified with the fragment separator FRS in the standard achromatic mode. ``The nuclei $^{209}$Hg, $^{210}$Hg, $^{211}$Tl, $^{212}$Tl, $^{218}$Bi, $^{219}$Po and $^{220}$Po have been identified for the first time.''

\subsection*{$^{221,222}$Po}
In the 2010 paper ``Discovery and investigation of heavy neutron-rich isotopes with time-resolved Schottky spectrometry in the element range from thallium to actinium'', Chen et al.\ described the discovery of $^{221}$Po and $^{222}$Po \cite{2010Che01}. A beryllium target was bombarded with a 670~MeV/u $^{238}$U beam from the GSI heavy-ion synchrotron SIS and projectile fragments were separated with the fragment separator FRS. The masses and half-lives of $^{221}$Po and $^{222}$Po were measured with time-resolved Schottky Mass Spectrometry in the storage-cooler ring ESR. ``In this experiment the new isotopes of $^{236}$Ac, $^{224}$At, $^{221}$Po, $^{222}$Po, and $^{213}$Tl were discovered.'' The half-lives of 112$^{+58}_{-28}$~s and 145$^{+604}_{-66}$~s for $^{221}$Po and $^{222}$Po, respectively, were listed in a table and are currently the only measured values.

\subsection*{$^{223-227}$Po}
$^{223}$Po, $^{224}$Po, $^{225}$Po, $^{226}$Po, and $^{227}$Po were discovered by Alvarez-Pol and the results were published in the 2010 paper ``Production of new neutron-rich isotopes of heavy elements in fragmentation reactions of $^{238}$U projectiles at 1A GeV'' \cite{2010Alv01}. A beryllium  target was bombarded with a 1~A GeV $^{238}$U beam from the GSI SIS synchrotron. The isotopes were separated and identified with the high-resolving-power magnetic spectrometer FRS. ``To search for new heavy neutron-rich nuclei, we tuned the FRS magnets for centering the nuclei $^{227}$At, $^{229}$At, $^{216}$Pb, $^{219}$Pb, and $^{210}$Au along its central trajectory. Combining the signals recorded in these settings of the FRS and using the analysis technique previously explained, we were able to identify 40 new neutron-rich nuclei with atomic numbers between Z=78 and Z=87; $^{205}$Pt, $^{207-210}$Au, $^{211-216}$Hg, $^{214-217}$Tl, $^{215-220}$Pb, $^{219-224}$Bi, $^{223-227}$Po, $^{225-229}$At, $^{230,231}$Rn, and $^{233}$Fr.''


\section{Summary}
The discoveries of the known thallium, lead, bismuth, and polonium isotopes have been compiled and the methods of their production discussed. About 30 of these isotopes ($\sim$17\%) were initially identified incorrectly. Half-lives were measured incorrectly or assigned to the wrong isotopes for the following isotopes: $^{181,182,191,200,201,204}$Tl, $^{181,198,202,203,205}$Pb, $^{198,199,201,206,207,210}$Bi, and $^{192-199}$Po. In addition, $^{203}$Pb, $^{205}$Pb, and $^{207}$Pb had been incorrectly reported to be stable.

\ack

This work was supported by the National Science Foundation under grants No. PHY06-06007 (NSCL) and PHY10-62410 (REU).

\bibliography{../isotope-discovery-references}

\newpage

\newpage

\TableExplanation

\bigskip
\renewcommand{\arraystretch}{1.0}

\section{Table 1.\label{tbl1te} Discovery of thallium, lead, bismuth, and polonium isotopes }
\begin{tabular*}{0.95\textwidth}{@{}@{\extracolsep{\fill}}lp{5.5in}@{}}
\multicolumn{2}{p{0.95\textwidth}}{ }\\

Isotope & Thallium, lead, bismuth, and polonium isotope \\
First author & First author of refereed publication \\
Journal & Journal of publication \\
Ref. & Reference \\
Method & Production method used in the discovery: \\

  & FE: fusion evaporation \\
  & NC: Neutron capture reactions \\
  & LP: light-particle reactions (including neutrons) \\
  & MS: mass spectroscopy \\
  & RD: radioactive decay \\
  & AS: atomic spectroscopy \\
  & SP: spallation reactions \\
  & PF: projectile fragmentation of fission \\

Laboratory & Laboratory where the experiment was performed\\
Country & Country of laboratory\\
Year & Year of discovery \\
\end{tabular*}
\label{tableI}

\datatables 



\setlength{\LTleft}{0pt}
\setlength{\LTright}{0pt}


\setlength{\tabcolsep}{0.5\tabcolsep}

\renewcommand{\arraystretch}{1.0}

\footnotesize 

\begin{longtable}{@{\extracolsep\fill}llllllll@{}}
\caption{Discovery of thallium, lead, bismuth, and polonium. See page\ \pageref{tbl1te} for Explanation of Tables}
Isotope & First Author & Journal & Ref. & Method & Laboratory & Country & Year\\
\hline\\
\endfirsthead\\
\caption[]{(continued)}
Isotope & First author & Journal & Ref. & Method & Laboratory & Country & Year\\
\hline\\
\endhead
$^{176}$Tl & H. Kettunen & Phys. Rev. C &\cite{2004Ket01}& FE & Jyv\"askyl\"a & Finland &2004 \\
$^{177}$Tl & G.L. Poli & Phys. Rev. C &\cite{1999Pol01}& FE & Argonne & USA &1999 \\
$^{178}$Tl & M.P. Carpenter & Phys. Rev. Lett. &\cite{1997Car01}& FE & Argonne & USA &1997 \\
$^{179}$Tl & J.R.H. Schneider & Z. Phys. A &\cite{1983Sch01}& FE & Darmstadt & Germany &1983 \\
$^{180}$Tl & Yu.A. Lazarev & Europhys. Lett. &\cite{1987Laz01}& FE & Dubna & Russia &1987 \\
$^{181}$Tl & K.S. Toth & Phys. Rev. C &\cite{1996Tot02}& FE & Argonne & USA &1996 \\
$^{182}$Tl & A. Bouldjedri & Z. Phys. A &\cite{1991Bou01}& FE & CERN & Switzerland &1991 \\
$^{183}$Tl & U.J. Schrewe & Phys. Lett. B &\cite{1980Sch01}& FE & Darmstadt & Germany &1980 \\
$^{184}$Tl & K.S. Toth & Phys. Lett. B &\cite{1976Tot01}& FE & Oak Ridge & USA &1976 \\
$^{185}$Tl & K.S. Toth & Phys. Lett. B &\cite{1976Tot01}& FE & Oak Ridge & USA &1976 \\
$^{186}$Tl & J.H. Hamilton & Phys. Rev. Lett. &\cite{1975Ham01}& FE & Oak Ridge & USA &1975 \\
$^{187}$Tl & K.S. Toth & Phys. Lett. B &\cite{1976Tot01}& FE & Oak Ridge & USA &1976 \\
$^{188}$Tl & J. Vandlik & Bull. Acad. Sci. USSR &\cite{1970Van01}& SP & Dubna & Russia &1970 \\
$^{189}$Tl & T.B. Vandlik & JETP Lett. &\cite{1972Van01}& SP & Dubna & Russia &1972 \\
$^{190}$Tl & J. Vandlik & Bull. Acad. Sci. USSR &\cite{1970Van01}& SP & Dubna & Russia &1970 \\
$^{191}$Tl & J. Vandlik & Bull. Acad. Sci. USSR &\cite{1974Van01}& SP & Dubna & Russia &1974 \\
$^{192}$Tl & G. Andersson & J. Inorg. Nucl. Chem. &\cite{1961And01}& LP & Uppsala & Sweden &1961 \\
$^{193}$Tl & K.F. Chackett & J. Inorg. Nucl. Chem. &\cite{1960Cha01}& FE & Birmingham & UK &1960 \\
$^{194}$Tl & B. Jung & Nucl. Phys. &\cite{1960Jun01}& LP & Uppsala & Sweden &1960 \\
$^{195}$Tl & J.D. Knight & Phys. Rev. &\cite{1955Kni01}& LP & Brookhaven & USA &1955 \\
$^{196}$Tl & G. Andersson & Phil. Mag. &\cite{1955And01}& LP & Uppsala & Sweden &1955 \\
$^{197}$Tl & G. Andersson & Phil. Mag. &\cite{1955And01}& LP & Uppsala & Sweden &1955 \\
$^{198}$Tl & D.A. Orth & Phys. Rev. &\cite{1949Ort01}& LP & Berkeley & USA &1949 \\
$^{199}$Tl & D.A. Orth & Phys. Rev. &\cite{1949Ort01}& LP & Berkeley & USA &1949 \\
$^{200}$Tl & D.A. Orth & Phys. Rev. &\cite{1949Ort01}& LP & Berkeley & USA &1949 \\
$^{201}$Tl & H.M. Neumann & Phys. Rev. &\cite{1950Neu01}& LP & Berkeley & USA &1950 \\
$^{202}$Tl & R.S. Krishnan & Proc. Camb. Phil. Soc. &\cite{1940Kri01}& LP & Cambridge & UK &1940 \\
$^{203}$Tl & F.W. Aston & Nature &\cite{1931Ast02}& MS & Cambridge & UK &1931 \\
$^{204}$Tl & G. Harbottle & Phys. Rev. &\cite{1953Har02}& NC & Brookhaven & USA &1953 \\
$^{205}$Tl & F.W. Aston & Nature &\cite{1931Ast02}& MS & Cambridge & UK &1931 \\
$^{206}$Tl & P. Preiswerk & Compt. Rend. Acad. Sci. &\cite{1935Pre01}& NC & Paris & France &1935 \\
$^{207}$Tl & O. Hahn & Phys. Z. &\cite{1908Hah02}& RD & Berlin & Germany &1908 \\
$^{208}$Tl & O. Hahn & Verh. Deutsch. Phys. Gesell. &\cite{1909Hah02}& RD & Berlin & Germany &1909 \\
$^{209}$Tl & F. Hagemann & Phys. Rev. &\cite{1950Hag01}& RD & Argonne & USA &1950 \\
$^{210}$Tl & O. Hahn & Phys. Z. &\cite{1909Hah01}& RD & Berlin & Germany &1909 \\
$^{211}$Tl & M. Pfuetzner & Phys. Lett. B &\cite{1998Pfu01}& PF & Darmstadt & Germany &1998 \\
$^{212}$Tl & M. Pfuetzner & Phys. Lett. B &\cite{1998Pfu01}& PF & Darmstadt & Germany &1998 \\
$^{213}$Tl & L. Chen & Phys. Lett. B &\cite{2010Che01}& PF & Darmstadt & Germany &2010 \\
$^{214}$Tl & H. Alvarez-Pol & Phys. Rev. C &\cite{2010Alv01}& PF & Darmstadt & Germany &2010 \\
$^{215}$Tl & H. Alvarez-Pol & Phys. Rev. C &\cite{2010Alv01}& PF & Darmstadt & Germany &2010 \\
$^{216}$Tl & H. Alvarez-Pol & Phys. Rev. C &\cite{2010Alv01}& PF & Darmstadt & Germany &2010 \\
$^{217}$Tl & H. Alvarez-Pol & Phys. Rev. C &\cite{2010Alv01}& PF & Darmstadt & Germany &2010 \\
 & & & & & &  \\
 & & & & & &  \\
$^{179}$Pb & A.N. Andreyev& J. Phys. G &\cite{2010And01}& FE & Darmstadt & Germany &2010 \\
$^{180}$Pb & K.S. Toth & Z. Phys. A &\cite{1996Tot01}& FE & Berkeley & USA &1996 \\
$^{181}$Pb & K.S. Toth & Nucl. Phys. A &\cite{1989Tot01}& FE & Berkeley & USA &1989 \\
$^{182}$Pb & J.G. Keller & Nucl. Phys. A &\cite{1986Kel01}& FE & Darmstadt & Germany &1986 \\
$^{183}$Pb & U.J. Schrewe & Phys. Lett. B &\cite{1980Sch01}& FE & Darmstadt & Germany &1980 \\
$^{184}$Pb & J.P. Dufour & Z. Phys. A &\cite{1980Duf01}& FE & Orsay & France &1980 \\
$^{185}$Pb & C. Cabot & Nucl. Phys. A &\cite{1975Cab01}& FE & Orsay & France &1975 \\
$^{186}$Pb & H. Gauvin & Phys. Rev. Lett. &\cite{1972Gau01}& FE & Orsay & France &1972 \\
$^{187}$Pb & H. Gauvin & Phys. Rev. Lett. &\cite{1972Gau01}& FE & Orsay & France &1972 \\
$^{188}$Pb & H. Gauvin & Phys. Rev. Lett. &\cite{1972Gau01}& FE & Orsay & France &1972 \\
$^{189}$Pb & H. Gauvin & Phys. Rev. Lett. &\cite{1972Gau01}& FE & Orsay & France &1972 \\
$^{190}$Pb & H. Gauvin & Phys. Rev. Lett. &\cite{1972Gau01}& FE & Orsay & France &1972 \\
$^{191}$Pb & Y. Le Beyec & Phys. Rev. C &\cite{1974LeB01}& FE & Berkeley & USA &1974 \\
$^{192}$Pb & Y. Le Beyec & Phys. Rev. C &\cite{1974LeB01}& FE & Berkeley & USA &1974 \\
$^{193}$Pb & J.O. Newton & Nucl. Phys. A &\cite{1974New01}& FE & Berkeley & USA &1974 \\
$^{194}$Pb & B. Jung & Nucl. Phys. &\cite{1960Jun01}& LP & Uppsala & Sweden &1960 \\
$^{195}$Pb & G. Andersson & Ark. Fysik &\cite{1957And01}& LP & Uppsala & Sweden &1957 \\
$^{196}$Pb & G. Andersson & Ark. Fysik &\cite{1957And01}& LP & Uppsala & Sweden &1957 \\
$^{197}$Pb & G. Andersson & Phil. Mag. &\cite{1955And01}& LP & Uppsala & Sweden &1955 \\
$^{198}$Pb & G. Andersson & Phil. Mag. &\cite{1955And01}& LP & Uppsala & Sweden &1955 \\
$^{199}$Pb & H.M. Neumann & Phys. Rev. &\cite{1950Neu01}& LP & Berkeley & USA &1950 \\
$^{200}$Pb & H.M. Neumann & Phys. Rev. &\cite{1950Neu01}& LP & Berkeley & USA &1950 \\
$^{201}$Pb & H.M. Neumann & Phys. Rev. &\cite{1950Neu01}& LP & Berkeley & USA &1950 \\
$^{202}$Pb & D. Maeder & Phys. Rev. &\cite{1954Mae01}& LP & Amsterdam & Netherlands &1954 \\
$^{203}$Pb & W. Maurer & Z. Phys. &\cite{1942Mau01}& LP & Berlin & Germany &1942 \\
$^{204}$Pb & H. Schuler & Naturwiss. &\cite{1932Sch01}& AS & Berlin & Germany &1932 \\
$^{205}$Pb & J.R. Huizenga & Phys. Rev. &\cite{1954Hui01}& LP & Argonne & USA &1954 \\
$^{206}$Pb & F.W. Aston & Nature &\cite{1927Ast01}& MS & Cambridge & UK &1927 \\
$^{207}$Pb & F.W. Aston & Nature &\cite{1927Ast01}& MS & Cambridge & UK &1927 \\
$^{208}$Pb & F.W. Aston & Nature &\cite{1927Ast01}& MS & Cambridge & UK &1927 \\
$^{209}$Pb & R.S. Krishnan & Proc. Camb. Phil. Soc. &\cite{1940Kri01}& LP & Cambridge & UK &1940 \\
$^{210}$Pb & K.A. Hofmann& Ber. Deuts. Chem. Ges. &\cite{1900Hof01}& RD & Munich & Germany &1900 \\
$^{211}$Pb & A. Debierne & Compt. Rend. Acad. Sci. &\cite{1904Deb01}& RD & Paris & France &1904 \\
$^{212}$Pb & E. Rutherford & Phil. Trans. Roy. Soc. A &\cite{1905Rut01}& RD & McGill & Canada &1905 \\
$^{213}$Pb & F.D.S. Butement & J. Inorg. Nucl. Chem. &\cite{1964But02}& SP & Liverpool & UK &1964 \\
$^{214}$Pb & E. Rutherford & Phil. Mag. &\cite{1904Rut01}& RD & McGill & Canada &1904 \\
$^{215}$Pb & M. Pfuetzner & Phys. Lett. B &\cite{1998Pfu01}& PF & Darmstadt & Germany &1998 \\
$^{216}$Pb & H. Alvarez-Pol & Eur. Phys. J. A &\cite{2009Alv01}& PF & Darmstadt & Germany &2009 \\
$^{217}$Pb & H. Alvarez-Pol & Eur. Phys. J. A &\cite{2009Alv01}& PF & Darmstadt & Germany &2009 \\
$^{218}$Pb & H. Alvarez-Pol & Eur. Phys. J. A &\cite{2009Alv01}& PF & Darmstadt & Germany &2009 \\
$^{219}$Pb & H. Alvarez-Pol & Eur. Phys. J. A &\cite{2009Alv01}& PF & Darmstadt & Germany &2009 \\
$^{220}$Pb & H. Alvarez-Pol & Phys. Rev. C &\cite{2010Alv01}& PF & Darmstadt & Germany &2010 \\
 & & & & & &  \\
 & & & & & &  \\
$^{184}$Bi & A.N. Andreyev & Eur. Phys. J. A &\cite{2003And01}& FE & Darmstadt & Germany &2003 \\
$^{185}$Bi & C.N. Davids & Phys. Rev. Lett. &\cite{1996Dav01}& FE & Argonne & USA &1996 \\
$^{186}$Bi & J.C. Batchelder & Z. Phys. A &\cite{1997Bat01}& FE & Argonne & USA &1997 \\
$^{187}$Bi & J.C. Batchelder & Eur. Phys. J. A &\cite{1999Bat01}& FE & Argonne & USA &1999 \\
$^{188}$Bi & U.J. Schrewe & Phys. Lett. B &\cite{1980Sch01}& FE & Darmstadt & Germany &1980 \\
$^{189}$Bi & H. Gauvin & Nucl. Phys. A &\cite{1973Gau01}& FE & Orsay & France &1973 \\
$^{190}$Bi & H. Gauvin & Phys. Rev. Lett. &\cite{1972Gau01}& FE & Orsay & France &1972 \\
$^{191}$Bi & H. Gauvin & Phys. Rev. Lett. &\cite{1972Gau01}& FE & Orsay & France &1972 \\
$^{192}$Bi & N.I. Tarantin & Sov. J. Nucl. Phys. &\cite{1971Tar01}& FE & Dubna & Russia &1971 \\
$^{193}$Bi & N.I. Tarantin & Sov. J. Nucl. Phys. &\cite{1971Tar01}& FE & Dubna & Russia &1971 \\
$^{194}$Bi & N.I. Tarantin & Sov. J. Nucl. Phys. &\cite{1971Tar01}& FE & Dubna & Russia &1971 \\
$^{195}$Bi & N.I. Tarantin & Sov. J. Nucl. Phys. &\cite{1971Tar01}& FE & Dubna & Russia &1971 \\
$^{196}$Bi & S. Chojnacki & Acta Phys. Pol. B &\cite{1976Cho01}& FE & Dubna & Russia &1976 \\
$^{197}$Bi & N.I. Tarantin & Sov. J. Nucl. Phys. &\cite{1971Tar01}& FE & Dubna & Russia &1971 \\
$^{198}$Bi & H.M. Neumann & Phys. Rev. &\cite{1950Neu01}& LP & Berkeley & USA &1950 \\
$^{199}$Bi & H.M. Neumann & Phys. Rev. &\cite{1950Neu01}& LP & Berkeley & USA &1950 \\
$^{200}$Bi & H.M. Neumann & Phys. Rev. &\cite{1950Neu01}& LP & Berkeley & USA &1950 \\
$^{201}$Bi & H.M. Neumann & Phys. Rev. &\cite{1950Neu01}& LP & Berkeley & USA &1950 \\
$^{202}$Bi & D.G. Karraker & Phys. Rev. &\cite{1951Kar01}& LP & Berkeley & USA &1951 \\
$^{203}$Bi & H.M. Neumann & Phys. Rev. &\cite{1950Neu01}& LP & Berkeley & USA &1950 \\
$^{204}$Bi & J.J. Howland & Phys. Rev. &\cite{1947How01}& LP & Berkeley & USA &1947 \\
$^{205}$Bi & D.G. Karraker & Phys. Rev. &\cite{1951Kar01}& LP & Berkeley & USA &1951 \\
$^{206}$Bi & J.J. Howland & Phys. Rev. &\cite{1947How01}& LP & Berkeley & USA &1947 \\
$^{207}$Bi & L.S. Germain & Phys. Rev. &\cite{1950Ger01}& LP & Berkeley & USA &1950 \\
$^{208}$Bi & J.A. Harvey & Can. J. Phys. &\cite{1953Har01}& LP & MIT & USA &1953 \\
$^{209}$Bi & F.W. Aston & Nature &\cite{1924Ast03}& MS & Cambridge & UK &1924 \\
$^{210}$Bi & E. Rutherford & Nature &\cite{1905Rut02}& RD & McGill & Canada &1905 \\
$^{211}$Bi & E. Rutherford & Phil. Trans. Roy. Soc. A &\cite{1905Rut01}& RD & McGill & Canada &1905 \\
$^{212}$Bi & E. Rutherford & Phil. Trans. Roy. Soc. A &\cite{1905Rut01}& RD & McGill & Canada &1905 \\
$^{213}$Bi & F. Hagemann & Phys. Rev. &\cite{1947Hag01}& RD & Argonne & USA &1947 \\
$^{214}$Bi & E, Rutherford & Phil. Mag. &\cite{1904Rut01}& RD & McGill & Canada &1904 \\
$^{215}$Bi & E.K. Hyde & Phys. Rev. &\cite{1953Hyd01}& RD & Berkeley & USA &1953 \\
$^{216}$Bi & D.G. Burke & Z. Phys. A &\cite{1989Bur01}& SP & CERN & Switzerland &1989 \\
$^{217}$Bi & M. Pfuetzner & Phys. Lett. B &\cite{1998Pfu01}& PF & Darmstadt & Germany &1998 \\
$^{218}$Bi & M. Pfuetzner & Phys. Lett. B &\cite{1998Pfu01}& PF & Darmstadt & Germany &1998 \\
$^{219}$Bi & H. Alvarez-Pol & Eur. Phys. J. A &\cite{2009Alv01}& PF & Darmstadt & Germany &2009 \\
$^{220}$Bi & H. Alvarez-Pol & Eur. Phys. J. A &\cite{2009Alv01}& PF & Darmstadt & Germany &2009 \\
$^{221}$Bi & H. Alvarez-Pol & Eur. Phys. J. A &\cite{2009Alv01}& PF & Darmstadt & Germany &2009 \\
$^{222}$Bi & H. Alvarez-Pol & Eur. Phys. J. A &\cite{2009Alv01}& PF & Darmstadt & Germany &2009 \\
$^{223}$Bi & H. Alvarez-Pol & Eur. Phys. J. A &\cite{2009Alv01}& PF & Darmstadt & Germany &2009 \\
$^{224}$Bi & H. Alvarez-Pol & Phys. Rev. C &\cite{2010Alv01}& PF & Darmstadt & Germany &2010 \\
 & & & & & &  \\
 & & & & & &  \\
$^{186}$Po & A.N. Andreyev & Phys. Rev. C &\cite{2005And01}& FE & Darmstadt & Germany &2005 \\
$^{187}$Po & A.N. Andreyev & Phys. Rev. C &\cite{2005And01}& FE & Darmstadt & Germany &2005 \\
$^{188}$Po & A.N. Andreyev & Eur. Phys. J. A &\cite{1999And01}& FE & Darmstadt & Germany &1999 \\
$^{189}$Po & A.N. Andreyev & Eur. Phys. J. A &\cite{1999And01}& FE & Darmstadt & Germany &1999 \\
$^{190}$Po & J.C. Batchelder & Phys. Rev. C &\cite{1996Bat01}& FE & Berkeley & USA &1996 \\
$^{191}$Po & A.B. Quint & Z. Phys. A &\cite{1993Qui01}& FE & Darmstadt & Germany &1993 \\
$^{192}$Po & S. Della Negra & J. Phys. (Paris) Lett. &\cite{1977Del01}& FE & Orsay & France &1977 \\
$^{193}$Po & A. Siivola & Nucl. Phys. A &\cite{1967Sii01}& FE & Berkeley & USA &1967 \\
$^{194}$Po & A. Siivola & Nucl. Phys. A &\cite{1967Sii01}& FE & Berkeley & USA &1967 \\
$^{195}$Po & A. Siivola & Nucl. Phys. A &\cite{1967Sii01}& FE & Berkeley & USA &1967 \\
$^{196}$Po & A. Siivola & Nucl. Phys. A &\cite{1967Sii01}& FE & Berkeley & USA &1967 \\
$^{197}$Po & C. Brun & Phys. Lett. &\cite{1965Bru01}& LP & Orsay & France &1965 \\
$^{198}$Po & C. Brun & Phys. Lett. &\cite{1965Bru01}& LP & Orsay & France &1965 \\
$^{199}$Po & C. Brun & Phys. Lett. &\cite{1965Bru01}& LP & Orsay & France &1965 \\
$^{200}$Po & D.G. Karraker & Phys. Rev. &\cite{1951Kar02}& LP & Berkeley & USA &1951 \\
$^{201}$Po & D.G. Karraker & Phys. Rev. &\cite{1951Kar02}& LP & Berkeley & USA &1951 \\
$^{202}$Po & D.G. Karraker & Phys. Rev. &\cite{1951Kar01}& LP & Berkeley & USA &1951 \\
$^{203}$Po & D.G. Karraker & Phys. Rev. &\cite{1951Kar01}& LP & Berkeley & USA &1951 \\
$^{204}$Po & D.G. Karraker & Phys. Rev. &\cite{1951Kar01}& LP & Berkeley & USA &1951 \\
$^{205}$Po & D.G. Karraker & Phys. Rev. &\cite{1951Kar01}& LP & Berkeley & USA &1951 \\
$^{206}$Po & D.H. Templeton & Phys. Rev. &\cite{1947Tem01}& LP & Berkeley & USA &1947 \\
$^{207}$Po & D.H. Templeton & Phys. Rev. &\cite{1947Tem01}& LP & Berkeley & USA &1947 \\
$^{208}$Po & D.H. Templeton & Phys. Rev. &\cite{1947Tem01}& LP & Berkeley & USA &1947 \\
$^{209}$Po & E.L. Kelly & Phys. Rev. &\cite{1949Kel01}& LP & Berkeley & USA &1949 \\
$^{210}$Po & P. Curie & Compt. Rend. Acad. Sci. &\cite{1898Cur03}& RD & Paris & France &1898 \\
$^{211}$Po & E. Marsden & Nature &\cite{1913Mar01}& RD & Manchester & UK &1913 \\
$^{212}$Po & O. Hahn & Phys. Z. &\cite{1906Hah02}& RD & McGill & Canada &1906 \\
$^{213}$Po & F. Hagemann & Phys. Rev. &\cite{1947Hag01}& RD & Argonne & USA &1947 \\
$^{214}$Po & K. Fajans & Phys. Z. &\cite{1912Faj01}& RD & Karlsruhe & Germany &1912 \\
$^{215}$Po & H. Geiger & Phil. Mag. &\cite{1911Gei01}& RD & Manchester & UK &1911 \\
$^{216}$Po & H. Geiger & Phys. Z. &\cite{1910Gei02}& RD & Manchester & UK &1910 \\
$^{217}$Po & F.F. Momyer& Phys. Rev. &\cite{1956Mom01}& LP & Berkeley & USA &1956 \\
$^{218}$Po & E. Rutherford & Phil. Mag. &\cite{1904Rut01}& RD & McGill & Canada &1904 \\
$^{219}$Po & M. Pfuetzner & Phys. Lett. B &\cite{1998Pfu01}& PF & Darmstadt & Germany &1998 \\
$^{220}$Po & M. Pfuetzner & Phys. Lett. B &\cite{1998Pfu01}& PF & Darmstadt & Germany &1998 \\
$^{221}$Po & L. Chen & Phys. Lett. B &\cite{2010Che01}& PF & Darmstadt & Germany &2010 \\
$^{222}$Po & L. Chen & Phys. Lett. B &\cite{2010Che01}& PF & Darmstadt & Germany &2010 \\
$^{223}$Po & H. Alvarez-Pol & Phys. Rev. C &\cite{2010Alv01}& PF & Darmstadt & Germany &2010 \\
$^{224}$Po & H. Alvarez-Pol & Phys. Rev. C &\cite{2010Alv01}& PF & Darmstadt & Germany &2010 \\
$^{225}$Po & H. Alvarez-Pol & Phys. Rev. C &\cite{2010Alv01}& PF & Darmstadt & Germany &2010 \\
$^{226}$Po & H. Alvarez-Pol & Phys. Rev. C &\cite{2010Alv01}& PF & Darmstadt & Germany &2010 \\
$^{227}$Po & H. Alvarez-Pol & Phys. Rev. C &\cite{2010Alv01}& PF & Darmstadt & Germany &2010 \\
\end{longtable}

\end{document}